\documentclass[preprint,amsmath,amssymb,aip]{revtex4-1}

\usepackage{listings}       
\usepackage[colorlinks=true,linkcolor=blue]{hyperref} 
\usepackage{graphicx}       
\usepackage{epstopdf}       
\usepackage[utf8]{inputenc} 
\usepackage[T1]{fontenc}
\usepackage{mathptmx} 

\begin{document}

\preprint{JCP19-AR-05159}

\title[]{A flexible and adaptive grid algorithm for global optimization utilizing basin hopping Monte Carlo}

\author{Mart\'in Leandro Paleico}
\email{martin.paleico@uni-goettingen.de}
\affiliation{Universit\"{a}t G\"{o}ttingen, Institut f\"{u}r Physikalische Chemie, Theoretische Chemie, Tammannstra\ss{}e 6, 37077 G\"{o}ttingen, Germany}
\author{J\"org Behler}
\email{joerg.behler@uni-goettingen.de}
\affiliation{Universit\"{a}t G\"{o}ttingen, Institut f\"{u}r Physikalische Chemie, Theoretische Chemie, Tammannstra\ss{}e 6, 37077 G\"{o}ttingen, Germany}
\affiliation{International Center for Advanced Studies of Energy Conversion (ICASEC), Universit\"at G\"ottingen, Tammannstra\ss{}e 6, 37077 G\"ottingen, Germany}
\date{\today}

\begin{abstract}
Global optimization is an active area of research in atomistic simulations, and many algorithms have been proposed to date. A prominent example is basin hopping Monte Carlo, which performs a modified Metropolis Monte Carlo search to explore the potential energy surface of the system of interest. These simulations can be very demanding due to the high-dimensional configurational search space.
The effective search space can be reduced by utilizing grids for the atomic positions, but at the cost of possibly biasing the results if fixed grids are employed. In this paper, we present a flexible grid algorithm for global optimization that allows to exploit the efficiency of grids without biasing the simulation outcome. The method is general and applicable to very heterogeneous systems, such as interfaces between two materials of different crystal structure or large clusters supported at surfaces. As a benchmark case, we demonstrate its performance for the well-known global optimization problem of Lennard-Jones clusters containing up to 100 particles. In spite of the simplicity of this model potential, Lennard-Jones clusters represent a challenging test case, since the global minima for some ``magic'' numbers of particles exhibit geometries that are very different from those of clusters with only a slightly different size.

\textbf{The following article has been submitted to the Journal of Chemical Physics. After it is published, it will be found at \url{https://aip.scitation.org/journal/jcp}}

\end{abstract}

\maketitle 
\bigskip

\section{Introduction}

Even for systems of moderate size, the configuration space is huge and too large to be explored systematically. However, only a small part of this configuration space is energetically relevant, and many systems in chemistry, physics and biology are found in -- or close to -- their local or even global minimum configurations. For instance, minerals naturally appear in their global minimum crystal structures or metastable in a few energetically low local minima. The same applies to molecules as well, such as proteins, where the folded configuration is known to be a very narrow and deep global minimum compared to a plethora of local minima of unfolded configurations~\cite{P4052}. Consequently, identifying those configurations, e.g. through computational methods, is of vital importance for any meaningful further investigation~\cite{pillardy_recent_2001}. 

Global optimization (GO)~\cite{P4232,P5606,hartke_global_2011} aims at finding the global minimum (GM) of the multidimensional potential energy surface (PES) of a given system. For essentially all relevant systems it is  impossible to find this minimum analytically, not only because the analytic form of the PES is usually unavailable, but also because the complexity of configuration space grows quickly with the number of degrees of freedom (DOF) resulting in an overwhelming number of local minima. 
While one possible approach would be to attempt an enumeration and energy evaluation of all the generated candidates, such an approach is naturally limited to small systems with only a few atoms or degrees of freedom~\cite{bone_exhaustive_1997, cyvin_staggered_1997, hu_protein_2006, pollock_scaffold_2008}. Instead, most GO algorithms explore the PES of the system until good candidates for global minima are found. They can be classified into two categories: restricting the available configuration space to decrease the number of explored configurations; or increasing the speed/efficiency at which the PES is explored.

The first group of algorithms is based on the use of fixed grids for the atomic positions~\cite{rahm_beyond_2017} in conjunction with Metropolis Monte Carlo simulations~\cite{metropolis_equation_1953}. The search is thus reduced from $3N_{\textrm{atoms}}$ continuum coordinates to the number of points in the grid. This can be applied to both, more or less regular crystals, where the grid represents occupied or vacant points on a lattice,~\cite{P5557} as well as to molecules, where the grid for a polymer or peptide, for example, is defined by the typical dihedral angles that need to be considered. The disadvantage of this approach is biasing the search by assuming that the relevant global minima are to be found within the restricted configuration space, and ignoring the remaining DOFs rendering this approach inapplicable to situations in which significant deviations from the underlying grid are important. 

To date, a few methods employing variable lattice sites have been reported~\cite{shao_dynamic_2004}, but these approaches often rely on evaluating the quality of the generated sites or optimizing the position of these sites by utilizing the same potential as for the optimized system. This is of course possible for a Lennard-Jones cluster, but becomes increasingly expensive for realistic force fields or \textit{ab-initio} calculations. Additionally, the paper by Shao \textit{et al.}~\cite{shao_dynamic_2004} presents some additional possible improvements, such as assuming that the new sites should be distributed on the surface of a sphere (which is a fair assumption for LJ clusters but does cannot necessarily be assumed for more complex potentials or interfaces like supported clusters), and neglecting sites \textit{inside} the optimized cluster. More recently, a method has been proposed by Yu \textit{et al.}~\cite{yu_unbiased_2019} that relies on dense grids and ``smart'' grid evaluation techniques to accelerate this process, but this has the drawback of requiring multiple evaluations of the grid sites with the optimizing potential.

In the second group of algorithms we find e.g. basin hopping Monte Carlo (BHMC)~\cite{wales_global_1997}, where transitions between basins, i.e., local minima, of the PES are accomplished by performing MC displacement trial moves, often on the whole system, followed by a geometry optimization, before the new structure is accepted or rejected by evaluating the Metropolis criterion. In this way the system is encouraged to leave its current potential energy surface (PES) basin by increasing the acceptance ratio of the MC approach. Similar to this procedure is simulated annealing (SA)~\cite{kirkpatrick_optimization_1983}, where the basin transition is achieved by performing successive steps of a molecular dynamics (MD) simulation at high temperature, giving the system enough kinetic energy to overcome potential energy barriers, followed by a slow cooling phase into a deep local or even the global minimum. Another related approach is minima hopping, in which short MD trajectories are followed by local structural optimizations~\cite{goedecker_minima_2004,P2201}.

Some algorithms even combine both approaches, such as evolutionary and genetic algorithms (GA)~\cite{P1518,deaven_molecular_1995}. Here the dimensionality reduction is accomplished by codifying the system's degrees of freedom into a ``genetic code'' (in the case of genetic algorithms) or working directly in coordinate space (in the case of evolutionary algorithms), and generating new candidates in part by performing mutations in that subspace. Changes in the atomic configuration arise from these mutations and from the combination of the ``fittest'' candidates from each generation to generate novel combinations finally yielding the chemically relevant local minima. GAs have found many successful applications, e.g. in crystal structure prediction~\cite{P0862}, optimization of free and supported atomic clusters~\cite{vilhelmsen_systematic_2012, vilhelmsen_identification_2014, huang_structural_2019, buendia_study_2017, heydariyan_new_2018}, and proteins~\cite{bozkurt_genetic_2018}.

In this paper we propose an improved global optimization algorithm based on the use of a flexible grid, which allows for a significant reduction of the effective search space while the identification of new atomic configurations is not restricted by intrinsic assumptions about the structure of the grid. Instead, the grid evolves dynamically along with the visited configurations. In combination with BHMC this grid enables an efficient exploration of the PES to yield the global minimum. In Section~\ref{sec:methods}, after a short summary of the employed model potential and BHMC, we give an explanation of the algorithm and its implementation. The success of the algorithm under a number of different initial parameter conditions is tested in Section \ref{sec:results} along with its performance for the well-studied benchmark case of Lennard-Jones clusters~\cite{wales_global_1997,P3937} containing up to 100 atoms. We demonstrate that for all cluster sizes, the new algorithm is able to find the respective global minima. We notice that even the usually hard to find clusters formed by 38 and 75-77 particles and an additional one containing 98 atoms can be identified without the need for further biasing. 

\section{Methods}\label{sec:methods}
\subsection{The Lennard-Jones Potential}

To test our algorithm, we perform global optimizations of clusters in vacuum using the Lennard-Jones (LJ) 12-6 pair potential~\cite{jones_j._e._determination_1924} to obtain the potential energy of the system,  

\begin{equation}
\label{eq:LJ}
E=\sum_{i=1}^{N_{\rm atoms}}\sum_{j>i}^{N_{\rm atoms}}4\epsilon \bigg[ \Big( \frac{\sigma}{r_{ij}} \Big) ^{12}- \Big( \frac{\sigma}{r_{ij}} \Big) ^6 \bigg]\quad
\end{equation}

\noindent where $\epsilon$ is the depth of the potential well, $\sigma$ is the distance at which the potential becomes zero, and $r_{ij}$ is the distance between atoms $i$ and $j$.

The LJ potential offers the advantages that it is fast to evaluate and easy to interpret, but in spite of its simple functional form it gives rise to a surprisingly complex behavior, not only for a small number of atoms -- e.g. regarding the structure of global minima, magic numbers resulting in very stable clusters, and the behavior of binary clusters -- but also for condensed phases, e.g. a complex phase diagram and multiple crystal structures \cite{P1978}. This makes it a challenging benchmark case for GO algorithms, while tests still can be carried out in a controlled and easily reproducible way at low computational costs. 

Another major advantage for the present work is that the global minima of LJ clusters containing up to 110 atoms are well known~\cite{wales_global_1997} facilitating the validation of the obtained results.
Nevertheless, several of the global minima of LJ clusters are notoriously hard to find~\cite{wales_global_1997}. The reason is that due to different geometric motifs compared to the close lying local minima naive GO approaches can get stuck in higher energy configurations, and also very narrow funnels can be present in configuration space~\cite{goedecker_minima_2004}, which can easily be missed by GO algorithms. 

\subsection{Basin Hopping Monte Carlo}

The basin hopping Monte Carlo algorithm~\cite{wales_global_1997} was proposed by Wales and Doye as a way of exploring the PES of systems with deep basins of attractions, such as clusters. A closely related procedure was earlier described by Li and Scheraga~\cite{li_monte_1987} aiming at the very different field of protein folding simulations. 

In cluster systems, conventional Monte Carlo (MC) is very inefficient due to a very low acceptance probability for each move, since most distortions out of the minimum are likely to end in high energy configurations with respect to the original basin resulting in a rejection by the Metropolis criterion. Consequently, a large number of MC steps is required to advance the system's configuration and to explore the PES. While raising the temperature of the MC algorithm leads to an increased acceptance ratio, now basins are only superficially visited and not explored in detail, possibly leading to missed minima.

\begin{figure}
    \centering
    \includegraphics[width=0.95\linewidth]{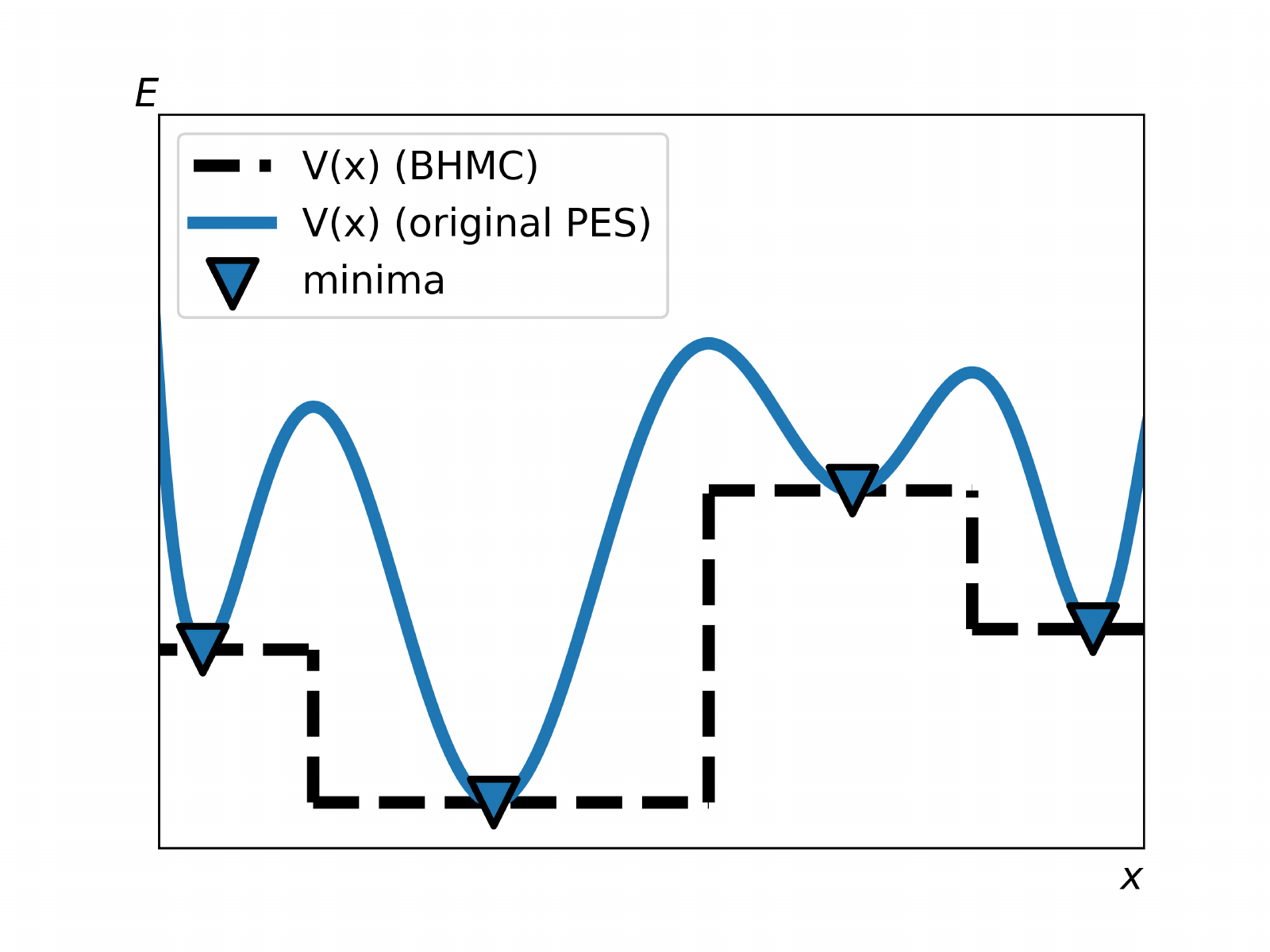}
    \caption{A one-dimensional PES as seen by the BHMC algorithm. The original PES $V(x)$ is transformed into a piece-wise step function representing the energy plateaus of the local minima generated by mapping every point to the corresponding minimum. All plots in this paper have been generated with the matplotlib library for python~\cite{hunter_matplotlib_2007}.} 
    \label{fig:bhmc}
\end{figure}

To avoid both of these problems, BHMC adds a local geometry optimization after each trial move. This relaxes every configuration to the closest local minimum such that effectively the whole PES is transformed to consist only of connected energy plateaus of the local minima (s. Fig.~\ref{fig:bhmc}), increasing the acceptance ratio of the algorithm and thus speeding up the exploration of the PES. This effect is so strong that in BHMC often it is possible to simultaneously displace \textit{all} the particles in the system in a trial move~\cite{wales_global_1997}, compared to typically only single particle displacement trial moves utilized in regular MC~\cite{frenkel_daan_understanding_2002}.

One limitation of the BHMC algorithm for some applications is that -- as originally designed -- it is restricted to ``local'' trial moves, which slows down the exploration of the PES by restricting jumps only to neighboring basins. More recent versions and modifications of the algorithm~\cite{P3937} have attempted to increase the type of possible moves by designing non-local trial moves tailored to cluster optimization, such as moving atoms from the inside to the outside of the cluster (as originally proposed by Takeuchi~\cite{takeuchi_clever_2006}). Still, the exploration of PESs by BHMC can be very demanding because of the high dimensionality of the configuration space. 

\subsection{The Flexible Grid Algorithm}
\subsubsection{Description of the Algorithm}

The use of grids in atomistic simulations has been proposed in many contexts and found numerous applications, from the investigation of alloys by the cluster expansion method~\cite{P2119} to the prediction of cluster shapes~\cite{rahm_beyond_2017}. In all these approaches, the employed grids are usually kept fixed corresponding in most cases to regular crystal structures. The grid points then define the possible atomic sites, which can be either occupied or vacancies, for which MC exchange moves can be performed. Note that fixed grids do not exclude structural relaxations, which are a crucial component e.g. of BHMC, but they define the possible initial atomic positions.

Utilizing a grid restricts the configuration space by discretizing the possible atomic positions, thus decreasing the number of configurations that need to be sampled by MC. Additionally, as the grid points represent physically meaningful lattice sites with their associated low potential energies, the acceptance probability of MC simulations can be substantially increased. However, fixed grids are not applicable to many important situations in which parts of the crystal exhibit strong structural changes, e.g. at grain boundaries~\cite{stukowski_visualization_2009} and for large-scale defects like screw dislocations, for crystal structure transitions, for amorphous systems, or generally if different materials are combined, like at interfaces or supported clusters~\cite{vilhelmsen_genetic_2014, vilhelmsen_identification_2014} (where a fixed grid becomes inconvenient if the surface can present defects, reconstructions, irregularly shaped adsorbates, etc.).
Utilizing a flexible grid avoids the biasing that is necessarily related to a fixed grid, and thus allows for a more general application also to these difficult but important scenarios.

In the flexible grid approach proposed here, grid points can be occupied (``active''), i.e., they contain a particle or atom relevant to the optimization problem under consideration, or unoccupied (``inactive''), i.e., they represent a vacancy or a target point for a possible exchange trial move. The grid resides inside a domain with or without periodic boundary conditions, which may also be just a subsystem of the full simulation setup.

Swaps are then attempted between occupied and unoccupied grid points, or in general also between sites occupied by different elements if present. All grid points are interacting, either by a physical force between the atoms if the grid points are occupied, or by a fictitious pair potential, if at least one of the two grid points in a given pair is unoccupied. The physical forces are determined by the chosen interaction potential and ensure a physically meaningful adjustment and relaxation of the atomic positions in the cluster. The purpose of the fictitious potential, which has no influence on the physical potential energy and the positions of the occupied grid points, is to adapt the positions of the empty grid points to provide optimum sites for future MC exchange moves.
When a BHMC swap has been successful, the empty grid points are relaxed around the new accepted atomic configuration, which itself is kept fixed after the atomic relaxation that is part of the BHMC step, thus allowing only unoccupied points to adapt to the new situation. In this way the grid structure flexibly follows the evolving structure of the system.

\begin{figure}
    \centering
    \includegraphics[width=0.95\linewidth]{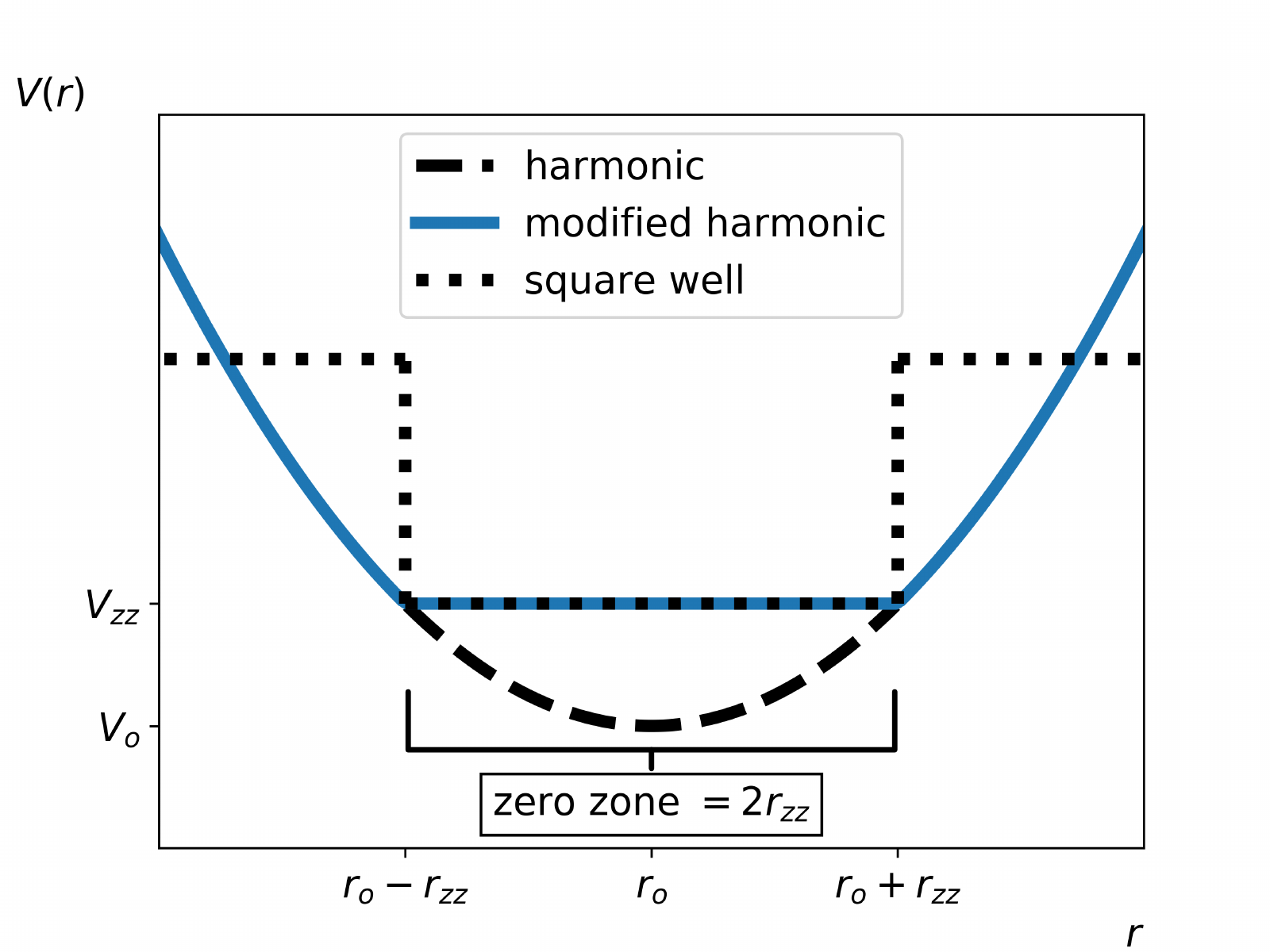}
    \caption{Fictitious interaction potential between the grid points (blue), compared to a purely harmonic potential (dashed black) and a square well potential (dotted black).}
    \label{fig:gridpotential}
\end{figure}

There are many possible choices for the fictitious potential connecting the grid points. Here we have decided to utilize a simple harmonic potential with a ``zero zone'' of constant potential in its central region, forming a softened square-well potential (s. Fig.~\ref{fig:gridpotential}). This choice has been motivated by the need to optimize systems where the nearest neighbor distance might not be known precisely \textit{a priori}, or where multiple nearest neighbor distances and/or bond types are present. Examples for such situations would be clusters where the equilibrium distance changes when switching from the bulk-like conditions in the center of the cluster towards the undercoordinated outermost shell, or amorphous carbon where single, double and triple bonds with their individual optimum equilibrium distances may be present. Using a flat region around the minimum of the potential, we avoid enforcing a single equilibrium distance for the whole grid, which would eventually bias the simulation outcome.

The ficticious potential is defined as

\begin{equation}
\label{eq:zero-zone-potential}
V(r)=\begin{cases}
k\cdot(r-r_0)^2      & \quad r< r_0-r_{zz}\vee r> r_0+r_{zz}  \\    
k\cdot(r_{zz}-r_0)^2 & \quad r_0-r_{zz}\le r \le r_0+r_{zz}   
\end{cases}
\end{equation}

\noindent where $k$ is the spring constant for the harmonic potential, $r_{0}$ is the equilibrium distance of the plain harmonic potential between grid points, and the constant $r_{zz}$ defines the width of the zero zone (between $r_0-r_{zz}$ and $r_0+r_{zz}$), where the potential has a constant value of $V_{zz}=k \cdot (r_{zz}-r_0)^2$. Note that since the potential in the zero zone is constant, its derivative is zero and thus no force is exerted between two grid points in this range, which allows to obtain any distance within the zero zone between the grid points in the optimization of the grid.

The fictitious potential serves to connect the points of the grid, suggesting some regularity and approximate equidistance in the grid, and concentrating grid points at reasonable distances to atomic positions where exchanges have a high probability of being accepted. Still, because of the zero zone of constant fictitious potential, the nearest neighbor distance for the atomic system does not need to be known precisely before the optimization starts. The shape of the harmonic potential as determined by $k$ is somewhat flexible, and is only required to enable gradient-based geometry optimization algorithms to minimize the fictitious potential energy of the grid. There is a discontinuity in the forces when switching from the harmonic branches to the constant potential in the center, but this could be remedied using a soft switching function between the two sections and does not pose a problem when optimizing the grid.

The grid points are only connected to other nearby grid points, which is achieved by restricting the maximum number of neighbors considered for each point. Still, this connectivity is not permanent, and as the grid is flexible, the nearest neighbors of a given grid point might change as the simulation progresses. For this reason, the neighbor list of each grid point is constructed not only at the very beginning of the simulation, but is updated after each grid optimization. The construction of this neighbor list proceeds as follows: with the help of a Verlet list\cite{frenkel_daan_understanding_2002}, the nearest neighbors to a given grid point are found. The distances to these grid points are calculated, and they are sorted from closest to farthest. Neighbors are assigned in this order, until the predefined maximum number of neighbors has been reached. During the minimization of the grid structure, only the points in the neighbor list of a grid interact with each other. Neighbors are always mutual, so if grid point A considers grid point B a neighbor, A is also in B's neighbor list.
The maximum allowed distance to count as a neighbor is also restricted, so that very distant points are not connected to avoid the grid collapsing or losing flexibility. 

The algorithm alternates between performing on the one hand exchange moves between occupied and unoccupied sites using BHMC including the usual structural optimization before applying the acceptance criterion, and on the other hand relaxing the unoccupied part of the grid utilizing the fictitious grid potential when a new atomic configuration has been accepted. Grid points behave differently according to their occupancy in this latter half of the algorithm: occupied points are fixed in space and are not allowed to move since the underlying, physical potential governing the atomic positions has already determined the position of these points, while unoccupied grid points relax around those positions. In this way the simulation is divided into two parts: the ``real'' simulation domain, where atoms interact with one another using their ``true'' potential, i.e., the force field or atomistic potential, and where the energy of the exchange moves is evaluated; and the grid domain, which contains only the unoccupied grid points thus excluding the relaxation of those points corresponding to occupied sites.

Specifically, for the global optimization of LJ clusters the algorithm proceeds as follows: The grid is initialized within the assigned grid domain, as unoccupied points on a simple cubic grid with slightly distorted positions. The grid is then relaxed, and random grid points are occupied by LJ atoms until the desired number of atoms has been reached. The atoms then undergo a geometry optimization, and the grid is minimized around this initial atomic configuration. From this point onward, the exchange loop begins: occupied and unoccupied sites are exchanged on the grid using BHMC for one atom at a time until a successful exchange has been performed. When this happens, the grid is re-optimized around the now occupied, fixed grid points. This repeats until the algorithm is stopped, e.g. by finding the global minimum, or by reaching a predetermined number of steps. The pseudocode for this procedure is shown in code block \ref{code:pseudocode}.

\begin{figure*}
    \centering
    \includegraphics[width=0.8\linewidth]{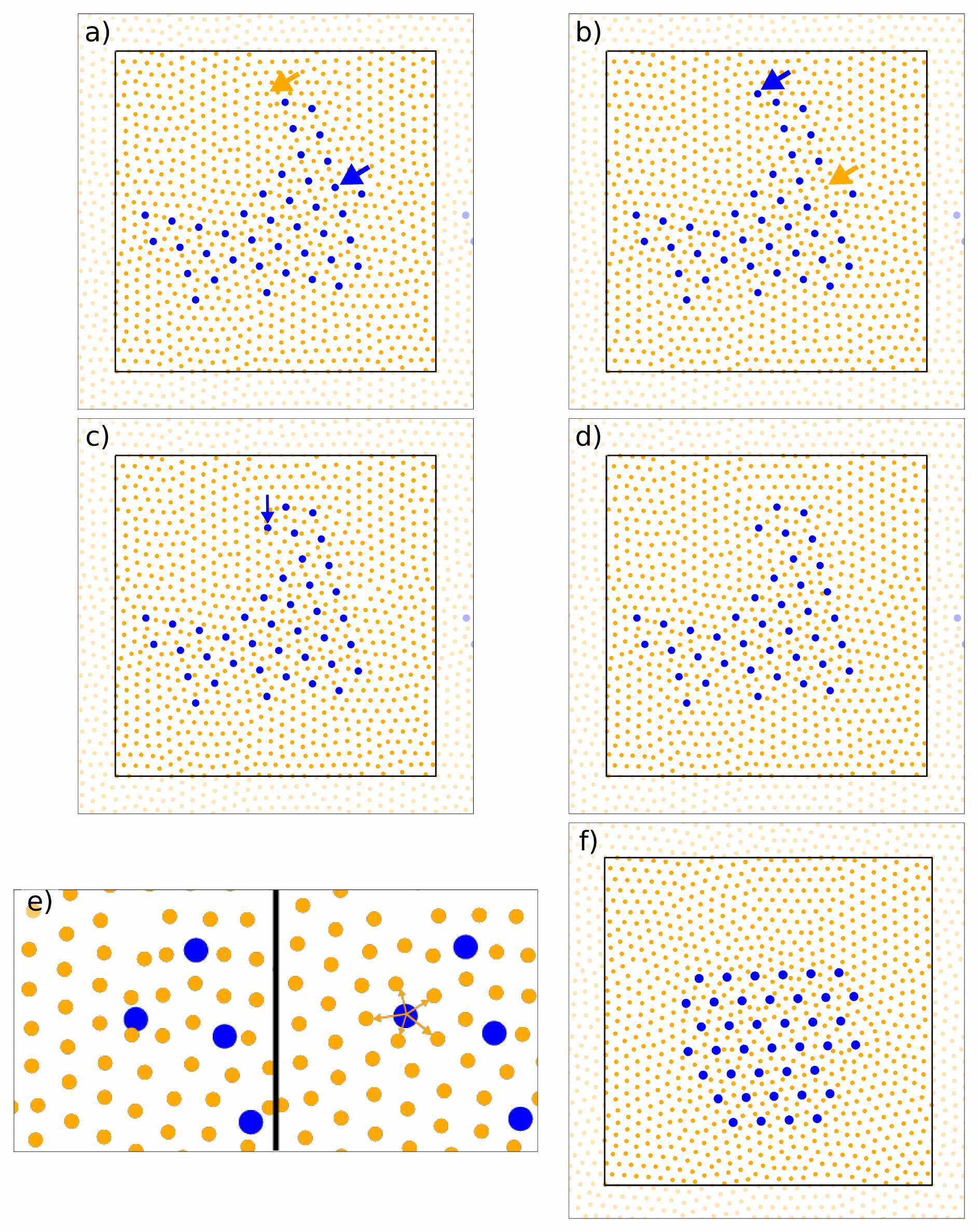}
    \caption{a) State of the grid (orange) and atoms (blue) before an exchange move. Periodic images of atoms across the periodic boundary conditions are shown slightly transparent. All atomistic images have been generated with OVITO~\cite{stukowski_visualization_2009}.
    b) State of the grid after an exchange move has been performed (circle around the exchanged positions). The new occupied site and the emerging empty grid point are highlighted.
    c) State of the system after the new cluster has been relaxed around the new exchanged positions using the physical potential, while the grid is still unchanged.
    d) State of the system after the grid has been relaxed around the new cluster configuration using the fictitious potential. Note that the grid has expanded around some positions.
    e) Zoomed in comparison of the grid pre (c) and post (d) relaxation.
    f) Final state of the system after 500 BHMC exchange steps.}
    \label{fig:2d-01-initial}
    \label{fig:2d-02-exchange}
    \label{fig:2d-03-relax}
    \label{fig:2d-04-relax-grid}
    \label{fig:2d-05-500steps}
\end{figure*}

For illustration purposes, Fig. \ref{fig:2d-01-initial}a to \ref{fig:2d-05-500steps}f show the procedure schematically for a 2D example. After the initialization of the grid and some BHMC steps, the system arrives in the configuration shown in Fig.~\ref{fig:2d-01-initial}a. From this configuration, an occupied and unoccupied site of the grid are chosen at random to perform an exchange resulting in the trial configuration shown in Fig.~\ref{fig:2d-02-exchange}b. Afterwards, the atoms of the cluster are relaxed (Fig.~\ref{fig:2d-03-relax}c), while the unoccupied part of the grid is still kept fixed. The cluster's energy is evaluated, and, assuming that the new potential energy is lower, the MC step is accepted. The grid now needs to relax around the new cluster positions (Fig.~\ref{fig:2d-04-relax-grid}d), which is also shown in more detail in Fig.~\ref{fig:2d-04-relax-grid}e. Repeating this procedure a total of 500 MC steps yields the configuration shown in Fig.~\ref{fig:2d-05-500steps}f, in which the cluster has now adopted a roughly hexagonal shape. A compressed xyz file containing the cluster particles and grid points of this trajectory is included in the supporting information, named grid-2d.xyz.

\begin{figure*}
    \centering
    \includegraphics[width=0.95\linewidth]{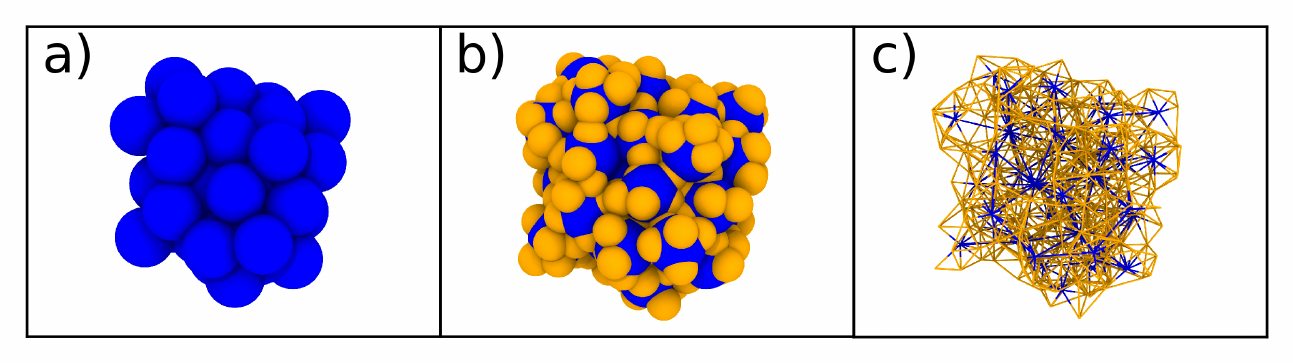}
    \caption{a) LJ cluster containing 38 atoms. b) Occupied grid points (i.e. the cluster) in blue and unoccupied grid points in orange. c) Connectivity of the cluster and the empty grid points. }
    \label{fig:3d-01-cluster}
    \label{fig:3d-02-clustergrid}
    \label{fig:3d-03-clustertubes}
\end{figure*}

For a real 3D case, Figure \ref{fig:3d-01-cluster}a shows a 38 LJ particle cluster (blue spheres). Figure \ref{fig:3d-02-clustergrid}b shows in addition the grid points within a 0.6 $\sigma$ radius of the cluster particles in orange. We can see how the grid arranges itself around the cluster atoms, but also fills in the space between them providing candidate vacancies for possible swapping events. Finally, Figure~\ref{fig:3d-03-clustertubes}c displays a bond-only/``wireframe'' view of Fig.~\ref{fig:3d-02-clustergrid}b, showing the link between close grid points and cluster atoms. A compressed xyz format file corresponding to the GO trajectory of this cluster including the grid points is available as supporting information, named grid-3d.xyz.

\begin{figure*} 
\begin{minipage}{0.7\linewidth}
\renewcommand{\lstlistingname}{Code Block}
\begin{lstlisting}[caption={Pseudocode for the flexible grid algorithm GO search for LJ clusters.},captionpos=b,label={code:pseudocode},numbers=left,frame=single, breaklines=true, postbreak=\mbox{\textcolor{red}{$\hookrightarrow$}\space},]
initialize grid
occupy grid positions with LJ particles
minimize LJ system
minimize grid
loop until (nsteps elapsed) or (known global minimum found)
    swap occupied and unoccupied grid position
    minimize LJ system
    evaluate Metropolis Monte Carlo criterion
    if accepted then
        check if known global minimum has been found
        minimize grid
    else if rejected then
        reset system to pre-swap state
end loop
\end{lstlisting}
\end{minipage}
\end{figure*}

\subsubsection{Parameters of the Grid}\label{sec:params_description}

The algorithm requires a number of input parameters that need to be specified before the GO search can start. The computational performance of the algorithm and the success ratio of the search depend on the values these parameters take. The optimal value ranges of the parameters may need to be adapted to the particular system under study, for instance because of very different typical interatomic distances. In this section, the relevant parameters are discussed, focusing on their meaning, their influence on the performance of the search, and how to estimate good initial values. Systematic tests will be reported in section~\ref{sec:param_sweep} and a summary is presented in table~\ref{tab:parameters}. These input parameters are: the multiplier for the number of grid points in the system ($N_{\textrm{mult}}$), the center point of the grid potential ($r_0$), the width of the zero zone (which depends on $r_{zz}$), and the maximum number of allowed neighbors for each grid point ($N_{\textrm{neigh}}$). Additionally, since the grid search is based on a BHMC search, the temperature ($T$) of the MC part of the algorithm needs to be specified in simulations employing the grid.

The number of grid points and their distribution in the grid domain should be adequate for the system of interest. In the case of global optimization of LJ clusters, we need enough grid points to cover the positions occupied by the LJ particles, and additional grid points on the surface of the cluster suggesting good positions outside of the cluster as well as points inside the clusters for swaps and vacancies. Points too far away from the cluster resulting in the formation of free atoms are poor candidates for swaps, and thus not needed. Evidently the number of grid points needs to increase as the number of atoms in the search does. Thus instead of defining an overall total number of grid points for each search and problem size, we define a multiplier $N_{\textrm{mult}}$ so that

\begin{equation}
    N_{\textrm{grid}} = N_{\textrm{LJ}} \cdot N_{\textrm{mult}}
\end{equation}

\noindent where $N_{\textrm{grid}}$ is the total number of grid points (both occupied and unoccupied),  $N_{\textrm{LJ}}$ is the number of Lennard-Jones particles in the simulation (or in general atoms to be optimized for other systems) and $N_{\textrm{mult}}$ is the multiplier. From this we can easily see that $\frac{1}{N_{\textrm{mult}}}$ defines the proportion of occupied sites in the grid.

This multiplier is one of the parameters having the strongest influence on the computational efficiency and scaling behavior of the algorithm. Using too few grid points would result in missing relevant swap positions and thus the GO search failing more frequently due to the coarse graining of space, while too many grid points will slow down the algorithm by offering irrelevant exchange positions away from the active sites that the BHMC still has to sample, and by slowing down the relaxation of the grid due to the need to calculate more interactions.

The estimated equilibrium distance between points in the grid is given by $r_0$, which should be short enough to locate grid points around atoms to allow for ``vacancies'' that can be occupied even inside the clusters. Too short spacings between grid points will, however result in physically repulsive structures upon occupation that can lead to long minimization times for the LJ clusters, or to numerical instabilities for some force fields and is computationally more demanding due to the increased number of grid points. A good starting estimate for this value is half of the nearest neighbor distance between atoms in the system of study.

These last two parameters automatically define a third quantity: the initial size of one side of the grid domain $L$ assuming for simplicity a cubic shape here. The relationship is given by

\begin{equation}
    L = \sqrt[3]{N_{\textrm{grid}}} 2 r_0 
\end{equation}

\noindent which assumes a cubic space around each grid point of size $2 r_0$.

The width of the zero zone given by $2r_{zz}$ is related to the flexibility of the grid. Without a zero zone, the  interaction potential of the grid would be fully harmonic yielding a single minimum, thus restricting substantially the number of optimized grid configurations. The introduction of the zero zone offers the advantage that it allows the nearest neighbor distance between grid points to adopt a range of values instead of a single value without an energy penalty. It allows grid points to lag behind, push, or drag other grid points near them, which is particularly important for points close to occupied grid points. If the value is too small -- or becomes zero in the fully harmonic case -- the grid configuration tends to get stuck, no longer being able to adapt to the changes in the real atom space. If the the width of the zero zone is too large, the grid points stop interacting and the grid becomes unstructured, once again losing its functionality. In the extreme case, when $r_0 = r_{zz}$, the grid potential loses its repulsive zone and becomes a flat plateau plus an attractive term, which must be avoided as too close grid points are unphysical.

The last relevant input parameter is the maximum number of allowed neighbors $N_{\textrm{neigh}}$ for each grid point. If the number of neighbors is too small, the grid cannot adopt a regular structure, and below $N_{\textrm{neigh}}=3$, it is not even able to fill 3D space. If $N_{\textrm{neigh}}$ is too large, it slows down computations since each pair of grid points requires one evaluation of the harmonic potential, and the system can once again get stuck as grid points are connected to other far away points which results in attractive forces in every direction. A good estimate for the proper value for this quantity is the coordination level usually found in solid systems or sphere packing, resulting in a 12-fold coordination (6-fold in 2D systems).

As a final parameter, as in any MC simulation, the temperature at which the search is performed needs to be specified. Too high search temperatures can result in leaving PES basins too easily, missing the chance to explore narrow PES funnels. On the other hand, if the search temperature is too low the algorithm can get stuck in basins, not being able to explore the rest of the PES. Additionally, the transition between the GM and close local minima might be hindered at some temperature ranges~\cite{wales_global_1997}.

We also find that some options in the algorithm either do not have a large influence on the GO searches (and thus are not required as input because they have safe default settings), or can be logically set without requiring tests in a range of values. The shape and periodicity of the domain in which the grid resides are more important, but depend clearly on the system under study. One should take into account the particularities (shape, symmetry breaking structural feature such as a cluster adsorbed on a surface) of the system. In this case, the grid domain could be restricted only to the part of the system of interest. 

The initial distribution of grid points in the simulation domain is not critical, since it will change and adapt as the algorithm progresses. Still, it is advantageous to avoid highly symmetrical initial conditions such as a perfectly cubic grid or a face-centered cubic configuration, that could bias the search by starting in a given PES basin. To avoid this, the algorithm by default initializes the grid atoms in a slightly, randomly distorted cubic mesh.

As for the potential between the grid points, the curvature of the harmonic potential $k$ only has a minor effect on the structure of the grid, as points of the adapted grids usually are located in the zero zones with constant minimum potential. This mostly has an influence on the behavior of the minimization algorithm. This value is set by default to 0.1~$\epsilon$.

The minimization algorithm and settings utilized by the grid is not important in our experience, as long as it is fast and can deal with the discontinuities in the derivatives of the grid potential. A simple gradient following, steepest descent algorithm was sufficient in our tests.

\subsubsection{Implementation}

The grid algorithm has been implemented in Python utilizing LAMMPS~\cite{P4473} as an external library responsible for performing the geometry optimization and energy evaluations of the LJ clusters, with a view towards utilizing LAMMPS for evaluation of more complex potentials in the future. 

We have used the default LAMMPS conjugate gradient method for local minimizations of the LJ clusters. For a 38 particle cluster and an energy convergence tolerance of 1 in $10^{-6}$, this requires an average of 70 minimization steps with a standard deviation of 40 steps. A simple BHMC test on the same cluster size following Wales and Doye's~\cite{wales_global_1997} settings but with the aforementioned local minimization parameters, required an average of 180 steps with a standard deviation of 55 steps. This difference is reasonable when we consider that the BHMC trial move involves all atoms in the cluster, but our grid swaps only involves atoms immediately in the vicinity of the origin and target sites.

The LAMMPS library could also be used to optimize the grid, but this is performed in Python in the current implementation, which at present is available for orthorhombic grid domains, while in principle a generalization to arbitrary domain shapes is possible. The source code is freely available under the GNU General Public License (GPL 3)~\cite{noauthor_gpl_nodate}, in a git repository: \url{https://gitlab.com/TheochemGoettingen/adaptive-grid}~.

\section{Results}\label{sec:results}
\subsection{Effect of grid parameters on the GO search} \label{sec:param_sweep}

\begin{figure}
    \centering
    \includegraphics[width=0.8\linewidth]{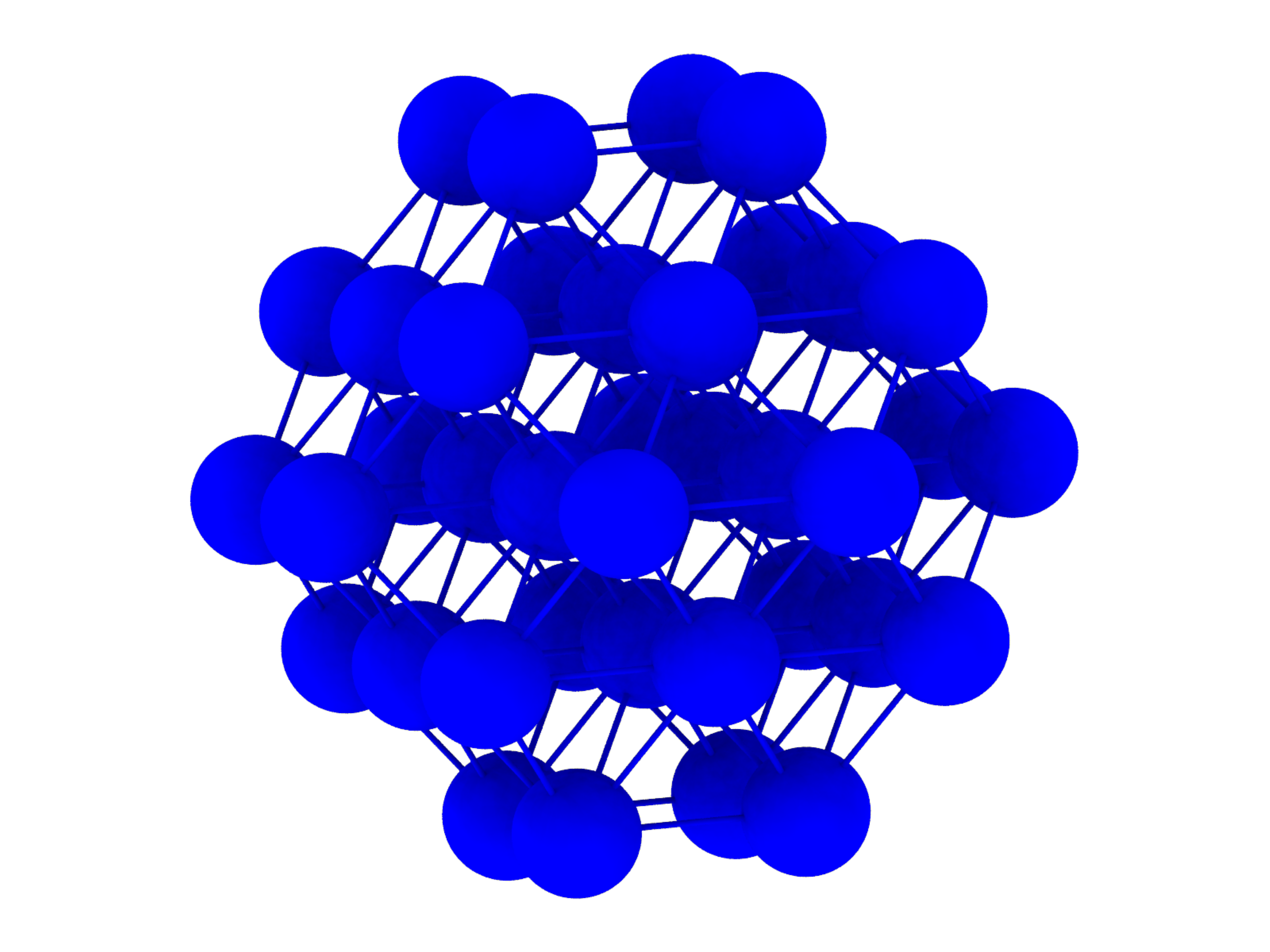}
    \caption{38 atom Lennard-Jones cluster utilized in testing the effect of different grid parameters on the effectiveness of the BHMC search.}
    \label{fig:ljcluster-38}
\end{figure}

To test the various parameters in the fictitious potential, we have chosen the 38 atom LJ cluster whose structure is shown in Fig.~\ref{fig:ljcluster-38}. We have made this choice as it represents one of the hardest to find global minima~\cite{wales_global_1997} of small-sized LJ clusters, but it is still found quickly enough to run repeated tests. The investigated parameters and their ranges of values are listed in Table~\ref{tab:parameters}.

\begin{figure*}
    \centering
    \includegraphics[width=0.95\linewidth]{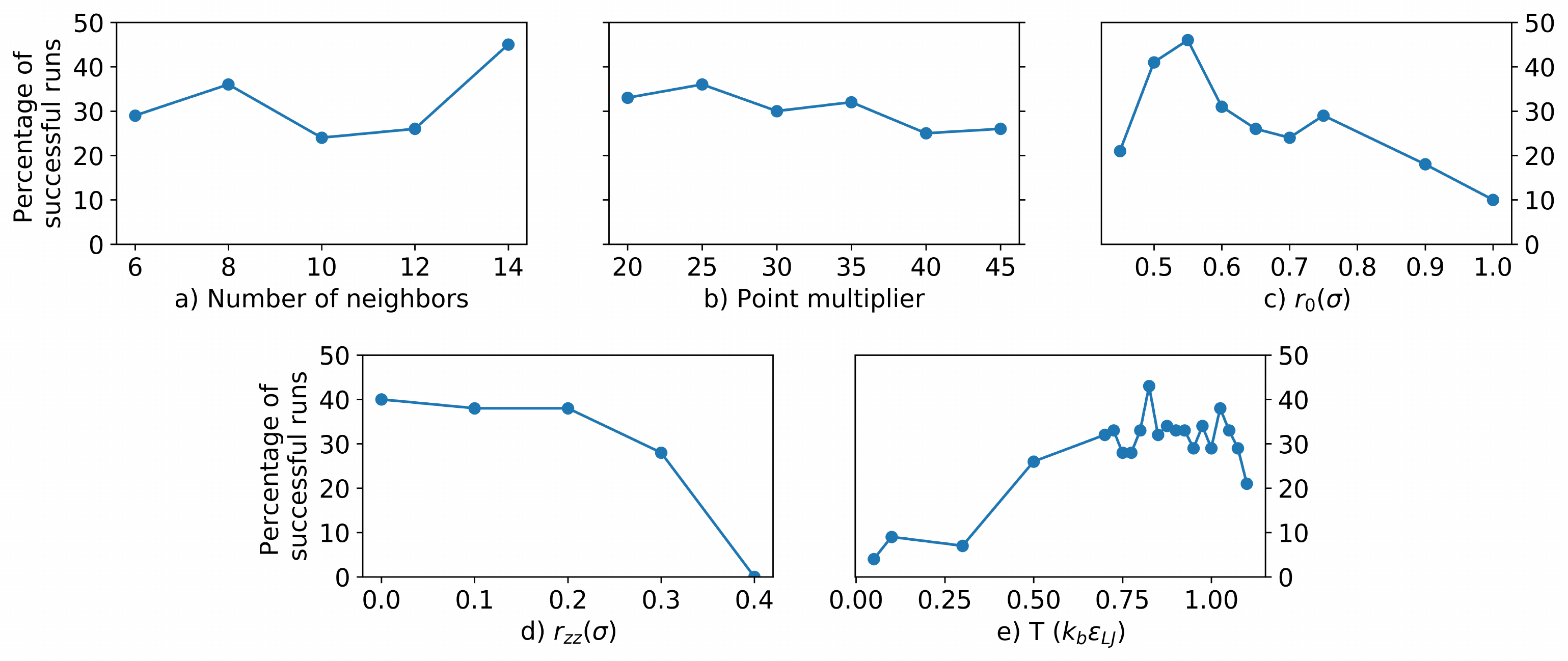}
    \caption{Optimal parameter search for a) the temperature ($T$)
    b) the grid point multiplier ($N_{\textrm{mult}}$)
    c) the center of the grid potential $r_0$ 
    d) the zero zone parameter ($r_{zz}$) and
    e) the number of neighbors ($N_{\textrm{neigh}}$) allowed for each grid point for a 38 atom LJ cluster.}
    \label{fig:param-temp}
    \label{fig:param-npoint}
    \label{fig:param-req}
    \label{fig:param-rzz}
    \label{fig:param-nneigh}
\end{figure*}

To investigate the role of these parameters, each GO search is run 100 times for a maximum of 10000 BHMC swap steps. For each parameter value, we track the percentage of successful runs. The results are presented in Figs. \ref{fig:param-temp}a to \ref{fig:param-temp}e. Table \ref{tab:parameters} shows the ranges tested for each parameter and the optimum values, while Table \ref{tab:constant-parameters} shows the values of all other parameters that are kept constant in the test while one parameter is being changed. 

\begin{table*}[t]
    \centering
    \begin{tabular}{|c|c|c|c|}
        \hline 
       Parameter  & Range  & Opt. Value & Meaning (units) \\
       \hline
        $T$                  & 0.7-1.1   & 0.8 & Temperature ($\epsilon_{LJ}/k_b$) \\
        $N_{\textrm{mult}}$  & 20-45     & 25    & Grid point number multiplier \\
        $r_0$                & 0.45-1.00 & 0.55  & Eq. distance of harmonic potential ($\sigma$) \\
        $r_{zz}$             & 0.0-0.4   & 0.0   & Half diameter of the zero zone ($\sigma$) \\
        $N_{\textrm{neigh}}$ & 8-14      & 14    & Number of grid neighbors \\
       \hline
    \end{tabular}
    \caption{Table of tested and optimal parameters for the grid algorithm}
    \label{tab:parameters}
\end{table*}

\begin{table}[t]
    \centering
    \begin{tabular}{|c|c|}
        \hline 
       Parameter (units) & Value \\
       \hline
        $T$ ($\epsilon_{LJ}/k_b$) & 0.825  \\
        $N_{\textrm{mult}}$       & 25  \\
        $r_0$ ($\sigma$)          & 0.6  \\
        $r_{zz}$ ($\sigma$)       & 0.1 \\
        $N_{\textrm{neigh}}$      & 12  \\
       \hline
    \end{tabular}
    \caption{Table of constant parameters for the parameter tests. If one parameter is being varied in the tests shown in Figs.~\ref{fig:param-temp}a to \ref{fig:param-temp}e, all other parameters are kept constant at the values given here.}
    \label{tab:constant-parameters}
\end{table}

For the number of neighbors in the grid (Fig. \ref{fig:param-nneigh}a) we see a somewhat flat probability curve that goes up for 14 allowed neighbors. This seems to indicate that over-coordination between the atomic positions and the grid helps the algorithm succeed. In any case, this is a test for this specific LJ system, and it might look different for other systems such as molecules.

In the case of the grid point multiplier $N_{\textrm{mult}}$ (Fig. \ref{fig:param-npoint}b), we observe a shallow maximum at 25, with the probability decreasing for smaller and larger multipliers. For smaller multipliers, this can be justified by not having enough points for the grid to cover the relevant configurations around the cluster. Higher multipliers instead add grid positions far away from the cluster (since the size of the simulation cell depends, among other things, on the number of grid points, see Section \ref{sec:params_description}), which are irrelevant for the GO search. Since the search is limited to a maximum number of steps, searches with a high multiplier waste steps on irrelevant configurations and terminate without reaching the GM.

For the parameters defining the fictitious potential $r_0$ and $r_{zz}$ (Figs. \ref{fig:param-req}c and \ref{fig:param-rzz}d respectively) we observe two different trends. For the equilibrium point of the grid we find two maxima, one at about 0.55~$\sigma$, and the other one at 0.75~$\sigma$. These correspond to about one half of the nearest ($2^{1/6} \sigma \approx 1.122 2^{1/6}$) and second nearest (about 1.55~$\sigma$ according to the radial distribution function of the clusters) neighbor distances for the LJ potential. These implies that the grid works at its optimum when the grid points position themselves halfway along the neighbor-neighbor distance. This allows the grid to fill any vacancies inside the cluster, which are good candidate positions for BHMC swaps.
For the zero zone size, we observe no or little decay in the success chance up to a half width of 0.2~$\sigma$, but a strong decline afterwards. In addition, the absence of the zero zone ($r_{zz}=0.0$) does not reduce the chances of finding the GM either. We conclude that for the simple LJ case with a well-defined minimum of the interparticle distance a constant potential zero zone is not needed, but it also does not significantly affect performance (up to a certain threshold value for the parameter) and might be of advantage in systems with more complex bonds and several differing bond distances.

Finally, for the temperature $T$ (Fig. \ref{fig:param-temp}e) we encounter an increasing probability of finding the GM as temperature increases, with a sharp increase at $T=0.4 \epsilon_{LJ}/k_b$. Beyond this value, the probability remains approximately constant across the sampled temperature range.

\subsection{GO Search}

To demonstrate the performance of the grid algorithm, we have searched for the global minima of all LJ cluster sizes between 4 and 100 particles. We find all the previously known GM for these sizes, as reported in the Cambridge Cluster Database~\cite{noauthor_cambridge_nodate}. For each cluster size up to 70 particles, we perform a flexible grid BHMC exchange search for at most 7000 steps, but terminate the search early if the known global minimum has been found. For clusters between 70 and 100 particles, we extend the search up to 15.000 steps. The search is repeated at least 25 times, but further searches are started if the global maximum has not yet been found, which was usually required for clusters larger than 60 particles. The algorithm requires only few steps to find the global minimum for sizes below 40 except for the well-known hard to find truncated octahedron with 38 particles, which required an average of 3500 steps, versus the 37 and 39 particle clusters at 2000 and 1500 average steps respectively. 

\begin{figure}
    \centering
    \includegraphics[width=0.95\linewidth]{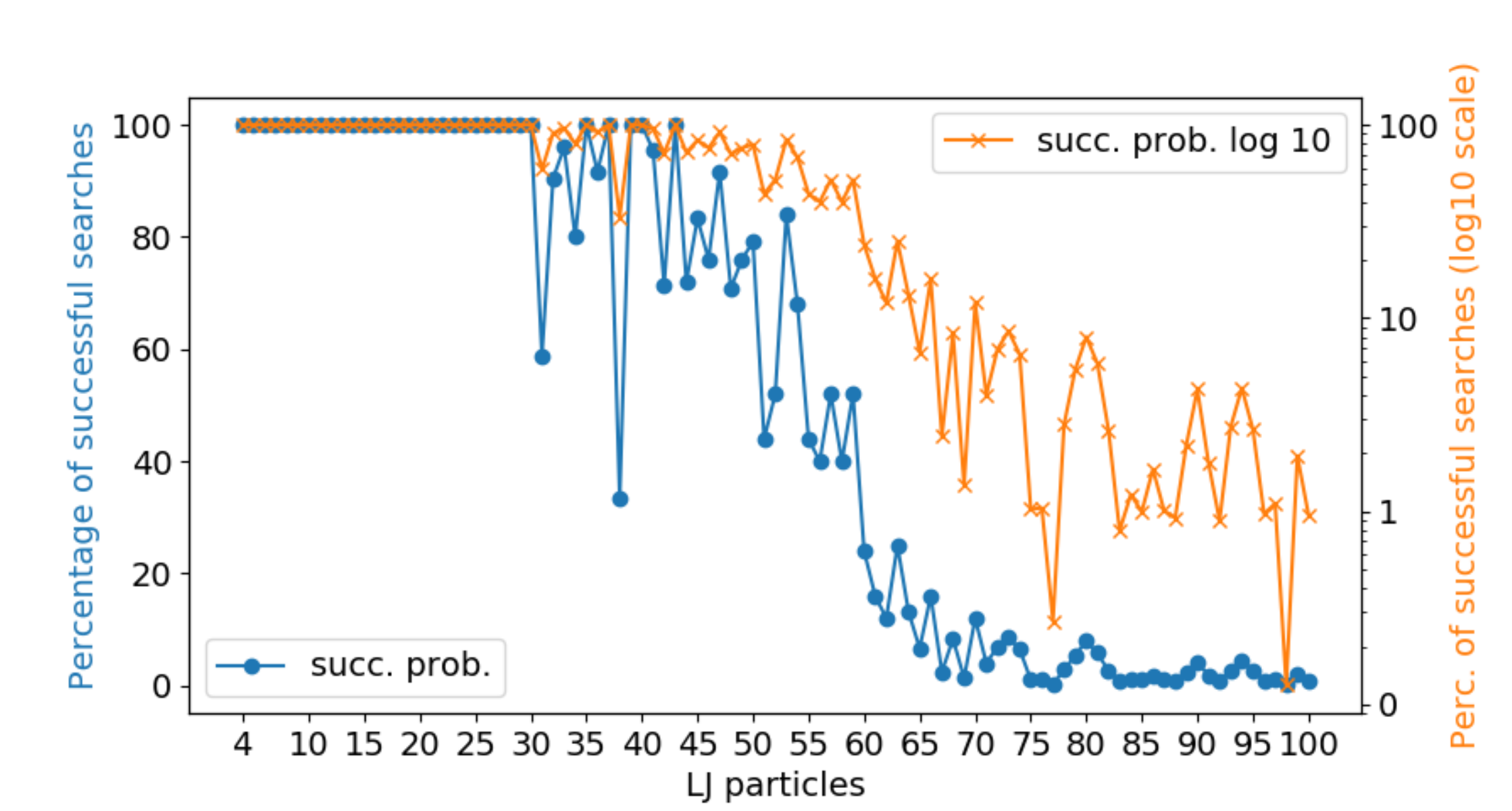}
    \caption{Percentage of successful GO search runs correctly identifying the global minimum structure as a function of the LJ cluster size. Right axis shows the same data, in base-10 logarithmic scale. Notice the valleys at 38, 75-77 and 98 atoms, corresponding to harder-to-find-than-average GMs.}
    \label{fig:percentage-succ}
    \label{fig:percentage-succ-log}
\end{figure}

\begin{figure}
    \centering
    \includegraphics[width=0.95\linewidth]{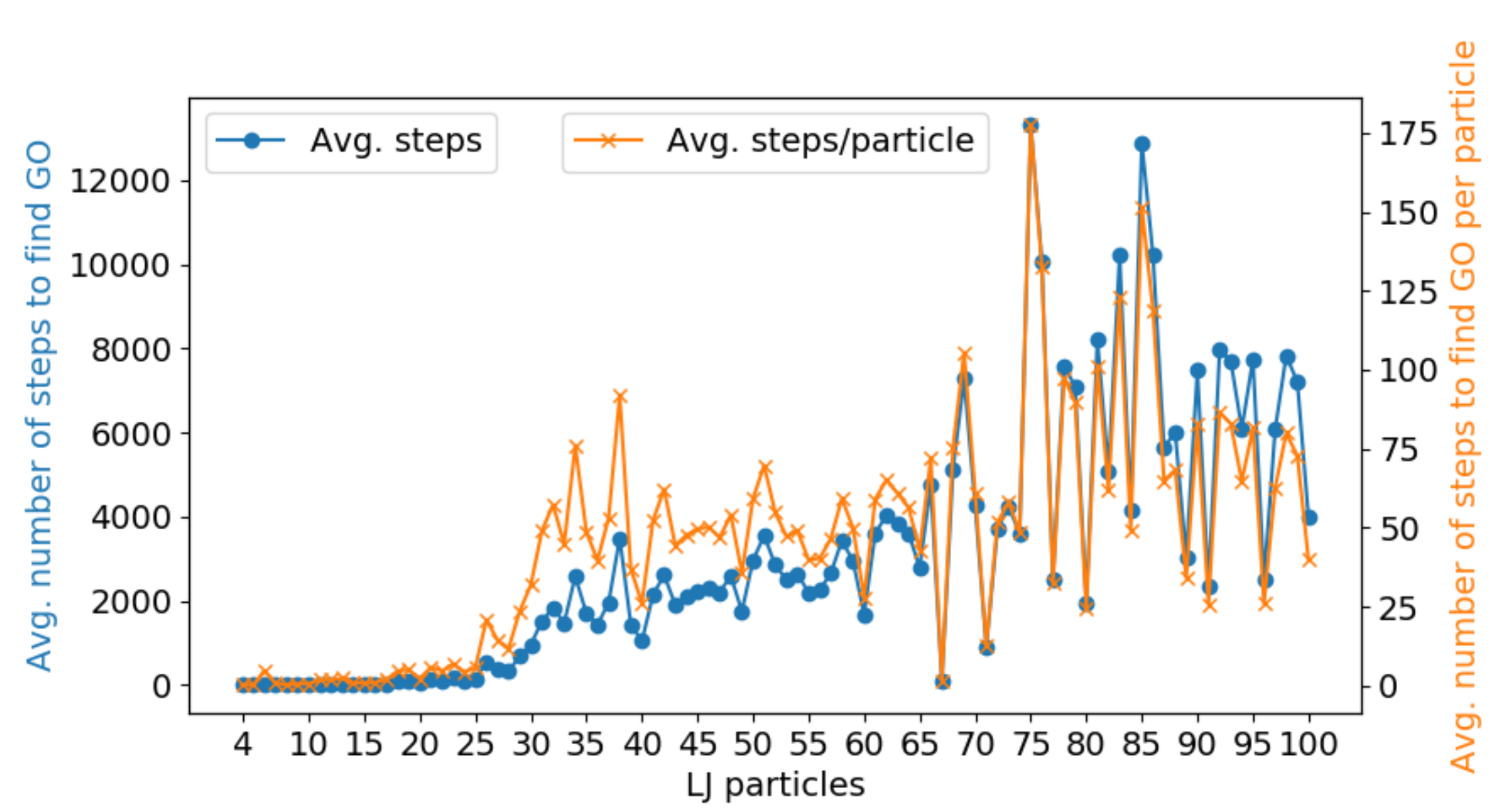}
    \caption{Average number of steps to first encounter the global minimum as a function of clusters size for simulations runs yielding the correct global minimum. The right axis shows the same data, weighted by the number of particles in each cluster. The fraction of successful runs is given in Figure~\ref{fig:percentage-succ}.}
    \label{fig:required-iterations}
    \label{fig:required-iterations-peratom}
\end{figure}

All known GM up to 60 particles have been found at least once in the first 25 runs. Beyond 60 particles, the landscape becomes much more complex, with the probability of finding the GM (Fig.~\ref{fig:percentage-succ}) decreasing sharply, and the average number of steps required to find this GM increasing rapidly (Fig.~\ref{fig:required-iterations}).

The difficulty of finding the GM increases more slowly when weighted by the number of particles in the system (as seen in Fig.~\ref{fig:required-iterations-peratom}), but still punctuated with cases that are particularly difficult (or easy) to find. Dividing by the number of atoms in the cluster allows for a closer comparison to the original BHMC search method, since in that algorithm all atoms are displaced in each step of the search, while here we only perform single atom moves by swapping a single occupied and an empty position in the grid.

It is interesting to note that for the hard to find cases, if the GM was found successfully, it was found within only a couple thousand of steps, although failed runs were allowed to continue for much longer. This reinforces the view that many GOs algorithms can get trapped in very narrow PES funnels: if a search is ``lucky'' enough to enter the correct funnel, the GM is quickly found. Otherwise, the search spend most of its time exploring other areas of the PES, belonging to wider basin local minima.

\begin{figure*}
    \centering
    \includegraphics[width=0.9\linewidth]{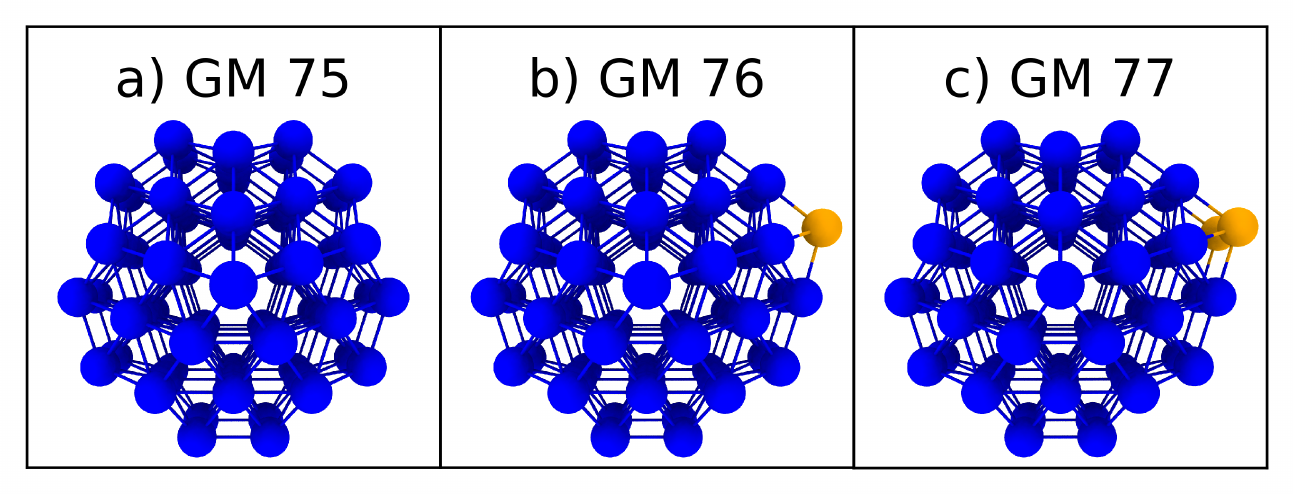}
    \caption{Global minima at a) 75, b) 76, and d) 77 LJ particles. Notice that they share a similar pentagonal basis shape (a), but 1 (b) and then 2 (c) atoms are added at a vertex (colored in orange).}
    \label{fig:hard-to-find-75-77}
\end{figure*}

In particular we want to highlight the hard to find minima between 75-77 particles (see Fig.~\ref{fig:hard-to-find-75-77}). At these particle numbers, local minima with icosahedral geometries are present~\cite{wales_global_1997}, and searches tend to repeatedly find these structures instead of the global minima. Note that we have used no pre-seeding procedure in our case. The proper GMs were found in each case independently as opposed to other reported BHMC searches~\cite{wales_global_1997} where seeding was required particularly for the 76 and 77 atom clusters of the 75-77 family.

\begin{figure*}
    \centering
    \includegraphics[width=0.6\linewidth]{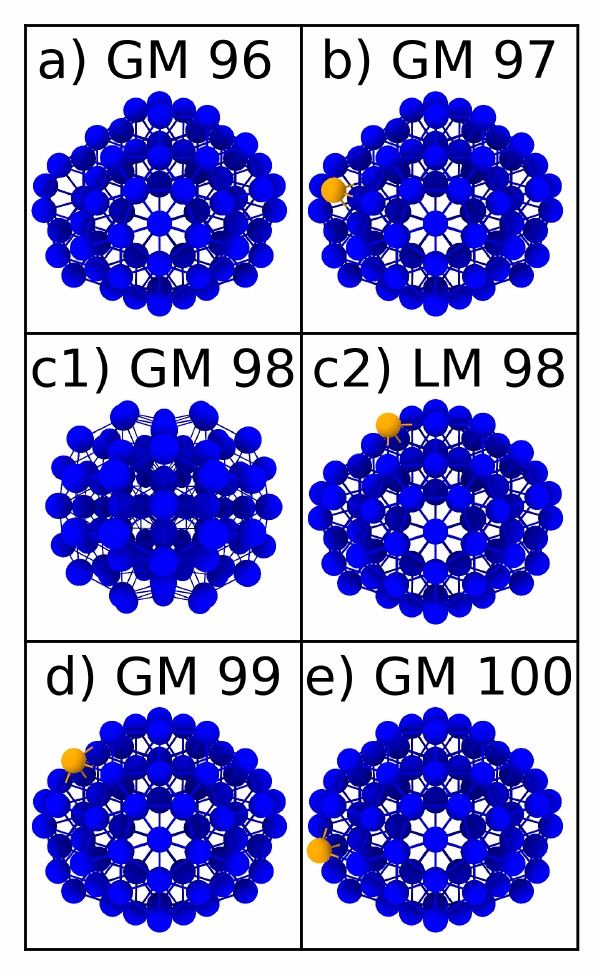}
    \caption{Global minima at a) 96, b) 97, c1) 98, d) 99 and e) 100 LJ particles, and the first local minimum at size 98 (c2). Notice that in c2), the local minimum is structurally related to the global minima at 96, 97, 99, and 100 in a) through d), while the GM c1) is not. Atoms in orange mark the added atoms in the structurally related series of clusters.}
    \label{fig:hard-to-find-97-99}
\end{figure*}

Another difficult to find GM is present for the cluster containing 98 particles. This cluster  exhibits a geometry differing significantly from its neighboring clusters at 97 and 99 particles (see Fig.~\ref{fig:hard-to-find-97-99}). Instead our algorithm often finds local minima similar in structure to the clusters at 97 and 99.

\subsection{Energy Profiles of the Grid Search}

\begin{figure}
    \centering
    \includegraphics[width=0.95\linewidth]{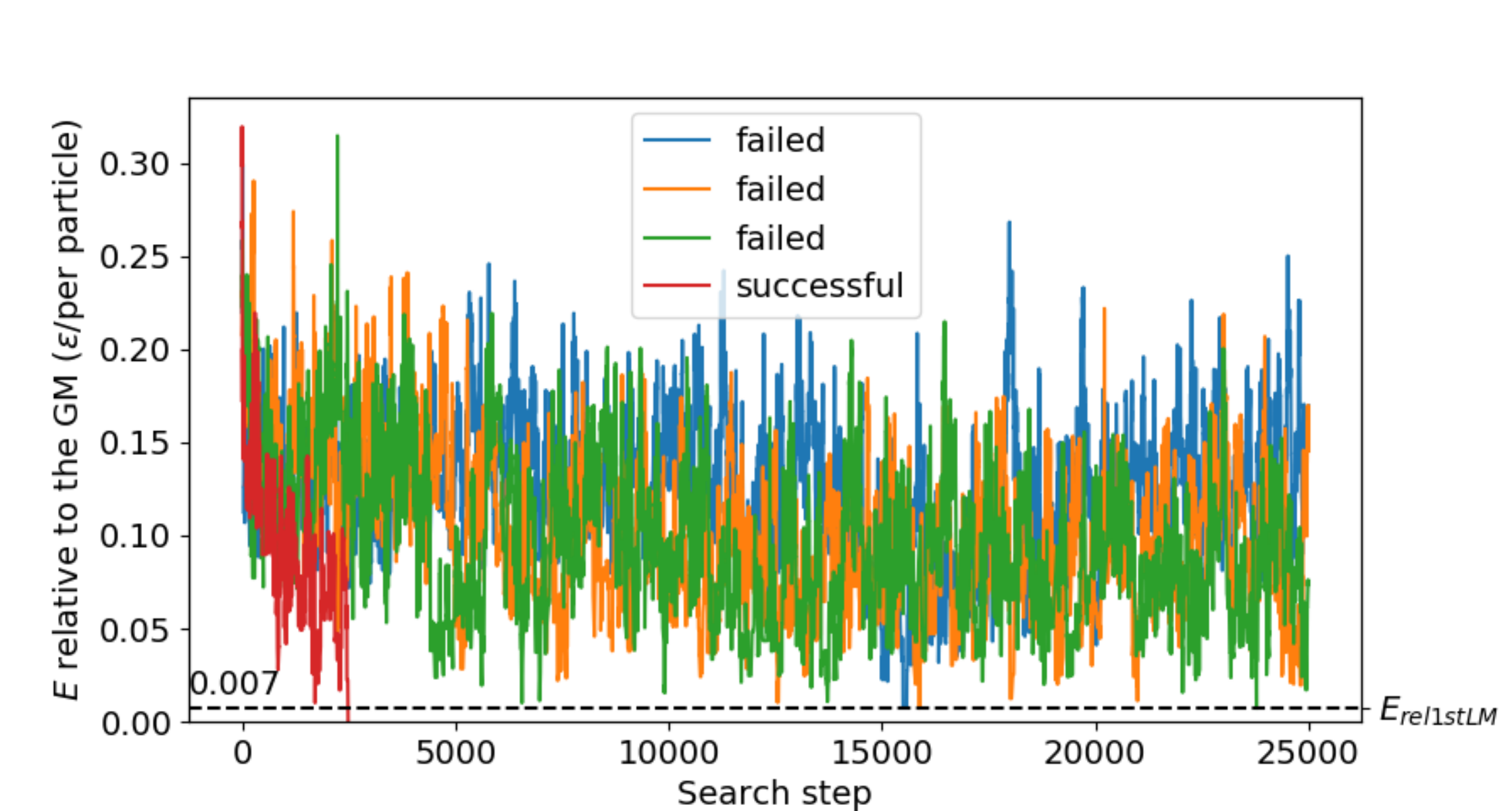}
    \caption{Energy profile for a grid search in the 77 particle LJ cluster, showing a succesful and three different failed runs. The dashed black line marks the relative energy per particle (0.007~$\epsilon$/particle above the GO) of the first icosahedral local minimum. Notice that the failed runs get close to the lowest energies of the successful run, but then increase in energy again, likely missing the funnel that leads to the GM.}
    \label{fig:energy-profile}
\end{figure}

It is instructive to compare failed and successful searches for certain hard to find cluster sizes. As an example, in Fig.~\ref{fig:energy-profile} we present four searches for the GM of the LJ cluster containing 77 particles. We observe that the successful run quickly goes down in energy, ends in the PES funnel leading to the GM, and identifies this minimum in only about 2500 steps. In contrast, the three failed runs only ever approach close energies much later in the respective runs. Instead these runs spend most of their time oscillating around icosahedral structures corresponding to local minima. 

\begin{figure}
    \centering
    \includegraphics[width=0.95\linewidth]{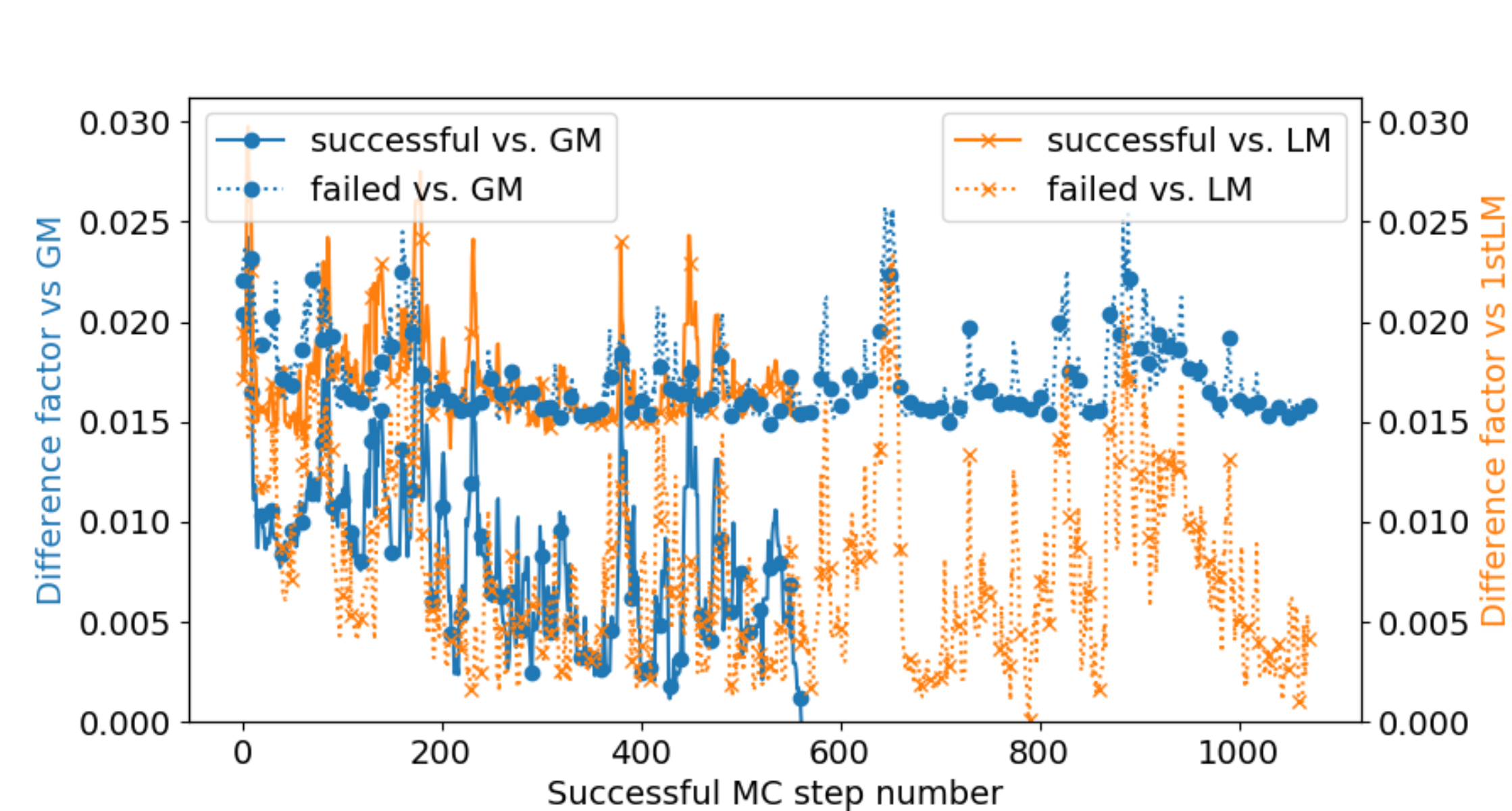}
    \caption{Difference factor vs. successful Monte Carlo step number for structures generated in a failed and successful GO search for the 77 particle cluster. The closer to zero the difference factor is, the more similar is the current structure to the GM or first LM, respectively. The successful run is displayed as a straight line, while failed run is plotted as dotted line; comparison vs. GM in blue with dot markers, vs. first LM in orange with cross markers.}
    \label{fig:struct-comp}
\end{figure}

This behavior is further analyzed by looking at the structural similarities between the current structure of the cluster and the known GM or the energetically lowest LM. This is plotted for one successful and one failed run in Fig.~\ref{fig:struct-comp}. Here the structures correspond to accepted BHMC steps. Shown is the difference factor (as implemented in the ASE~\cite{larsen_atomic_2017}, Atomic Simulation Environment Python library) based on interatomic distances within each cluster vs. the reference cluster~\cite{vilhelmsen_systematic_2012} for the structures visited in each search. The factor takes a value of 0.0 for an exact match, grows as the clusters become more different, and intrinsically takes into account rotations and translations about the center of mass of the system since it is based on interatomic distances. It can be seen that the successful run quickly ends in structural vicinity of the GM and stays there until it is finally adopted at around structure number 600, while the failed run stays in the funnel belonging to the first LM which is reached around structure number 800.

The trajectory corresponding to structures visited in Fig.~\ref{fig:struct-comp} are provided in the supporting information, under the names successful-77.xyz and failed-77.xyz.

\section{Conclusions}\label{sec:conclusions}

We have developed an adaptive grid algorithm for searching global minima of atomistic systems employing BHMC simulations. The algorithm is in principle unbiased with respect to the system's preferred crystal structure and other regular patterns, like packing, coordination or bond distances between atoms. The algorithm is thus very flexible, and ideal for situations where the underlying structural features are not known \textit{a priori}. 
We have demonstrated its applicability by identifying the global minima of LJ clusters containing up to 100 particles.
Since no particular lattice is assumed, a generalization of the procedure is straightforward to systems including multiple materials with arbitrary lattices. Examples of this are free-standing nanoparticles, or interfaces between clusters or nanostructures and solid surfaces.

The adaptive grid does introduce some overhead as it needs to be minimized after each successful BHMC step. This makes it ideal for medium-speed force fields like machine learning potentials~\cite{P4885}. Fast force fields would allow to perform a pure BHMC search without the grid's overhead by drastically increasing the number of trial steps, while very expensive potentials like an on-the-fly computation of the energy by electronic structure methods requires a more biased search approach to avoid wasting time on rejected structures. An advantage of our method when compared to other variable lattice approaches~\cite{shao_dynamic_2004, yu_unbiased_2019} is that our grid construction and optimization is decoupled from the underlying potential of the cluster. In this way, the cost for constructing the grid remains constant if more complex physical potentials are used.

A disadvantage of our approach when compared to conventional BHMC is that our trial moves affect only one atom at a time. Performing multiple swaps would lead to large jumps in PES space, leading the algorithm out of the current basin instead of exploring it extensively as with single swaps. To compensate for this, our approach can easily be combined with the regular local BHMC displacement moves, as well as improvements such as non-local moves~\cite{rondina_revised_2013} and occasional jumping~\cite{iwamatsu_basin_2004}, or the usual moves used in $NPT$-MC (volume changes) or $\mu VT$ grand canonical MC (deletion/addition of atoms, for which the grid itself could be used as a target).

The algorithm can also be adapted to other methods that do not perform global searches. For example, the equilibrium between clusters adsorbed on a surface could easily be modelled to study Ostwald ripening~\cite{voorhees_theory_1985} under different conditions such as monolayer coverage and simulation temperature. This sort of simulation is hard to perform without a grid~\cite{lucas_simulation_2010}, since many possible MC moves displace an atom into unfavorable positions, resulting in low acceptance ratios. Another target of interest is Grand Canonical Ensemble simulations, where atoms are adsorbed into a surface from an imaginary reservoir under different conditions of temperature and chemical potential/partial pressure. Once again, a grid facilitates the simulation by reducing the number of possible adsorption sites to a number of good candidates.

Another possible future extension would be the implementation of active sites~\cite{rahm_beyond_2017}. Here exchanges are only allowed between occupied sites, and active unoccupied sites within a certain distance of already occupied sites. This avoids wasting time in performing exchanges with unoccupied sites far away from the cluster, with minimal investment.

\section{Supplementary Material}

As described throughout the text, a number of files are provided as supplementary material:

\begin{enumerate}
    \item A compressed xyz file containing the cluster particles and grid points of the example 2D grid simulation, named grid-2d.xyz.
    
    \item A compressed xyz format file corresponding to the GO trajectory of a 38 LJ particle 3D cluster, named grid-3d.xyz.
    
    \item The trajectory corresponding to structures visited in Fig.~\ref{fig:struct-comp}, comparing successful and failed searches for the GM of the 77 LJ particle cluster, are provided in the supporting information, under the names successful-77.xyz and failed-77.xyz.
\end{enumerate}

\begin{acknowledgments}
We thank the Deutsche Forschungsgemeinschaft (DFG) for financial support (Be3264/10-1, project number 289217282 and INST186/1294-1 FUGG, project number 405832858). JB gratefully acknowledges a DFG Heisenberg professorship (Be3264/11-2, project number 329898176). We would also like to thank the North-German Supercomputing Alliance (HLRN) under project number nic00046 for computing time. 
\end{acknowledgments}

\bibliography{grid-paper-bibtex-20-01-28.bib}

\begin{thebibliography}{46}%
\makeatletter
\providecommand \@ifxundefined [1]{%
 \@ifx{#1\undefined}
}%
\providecommand \@ifnum [1]{%
 \ifnum #1\expandafter \@firstoftwo
 \else \expandafter \@secondoftwo
 \fi
}%
\providecommand \@ifx [1]{%
 \ifx #1\expandafter \@firstoftwo
 \else \expandafter \@secondoftwo
 \fi
}%
\providecommand \natexlab [1]{#1}%
\providecommand \enquote  [1]{``#1''}%
\providecommand \bibnamefont  [1]{#1}%
\providecommand \bibfnamefont [1]{#1}%
\providecommand \citenamefont [1]{#1}%
\providecommand \href@noop [0]{\@secondoftwo}%
\providecommand \href [0]{\begingroup \@sanitize@url \@href}%
\providecommand \@href[1]{\@@startlink{#1}\@@href}%
\providecommand \@@href[1]{\endgroup#1\@@endlink}%
\providecommand \@sanitize@url [0]{\catcode `\\12\catcode `\$12\catcode
  `\&12\catcode `\#12\catcode `\^12\catcode `\_12\catcode `\%12\relax}%
\providecommand \@@startlink[1]{}%
\providecommand \@@endlink[0]{}%
\providecommand \url  [0]{\begingroup\@sanitize@url \@url }%
\providecommand \@url [1]{\endgroup\@href {#1}{\urlprefix }}%
\providecommand \urlprefix  [0]{URL }%
\providecommand \Eprint [0]{\href }%
\providecommand \doibase [0]{http://dx.doi.org/}%
\providecommand \selectlanguage [0]{\@gobble}%
\providecommand \bibinfo  [0]{\@secondoftwo}%
\providecommand \bibfield  [0]{\@secondoftwo}%
\providecommand \translation [1]{[#1]}%
\providecommand \BibitemOpen [0]{}%
\providecommand \bibitemStop [0]{}%
\providecommand \bibitemNoStop [0]{.\EOS\space}%
\providecommand \EOS [0]{\spacefactor3000\relax}%
\providecommand \BibitemShut  [1]{\csname bibitem#1\endcsname}%
\let\auto@bib@innerbib\@empty
\bibitem [{\citenamefont {Lin}\ and\ \citenamefont {Zewail}(2012)}]{P4052}%
  \BibitemOpen
  \bibfield  {author} {\bibinfo {author} {\bibfnamefont {M.~M.}\ \bibnamefont
  {Lin}}\ and\ \bibinfo {author} {\bibfnamefont {A.~H.}\ \bibnamefont
  {Zewail}},\ }\bibfield  {title} {\enquote {\bibinfo {title} {Protein folding
  – simplicity in complexity},}\ }\href@noop {} {\bibfield  {journal}
  {\bibinfo  {journal} {Ann. Phys.}\ }\textbf {\bibinfo {volume} {524}},\
  \bibinfo {pages} {379--391} (\bibinfo {year} {2012})}\BibitemShut {NoStop}%
\bibitem [{\citenamefont {Pillardy}\ \emph {et~al.}(2001)\citenamefont
  {Pillardy}, \citenamefont {Czaplewski}, \citenamefont {Liwo}, \citenamefont
  {Lee}, \citenamefont {Ripoll}, \citenamefont {Kaźmierkiewicz}, \citenamefont
  {Ołdziej}, \citenamefont {Wedemeyer}, \citenamefont {Gibson}, \citenamefont
  {Arnautova}, \citenamefont {Saunders}, \citenamefont {Ye},\ and\
  \citenamefont {Scheraga}}]{pillardy_recent_2001}%
  \BibitemOpen
  \bibfield  {author} {\bibinfo {author} {\bibfnamefont {J.}~\bibnamefont
  {Pillardy}}, \bibinfo {author} {\bibfnamefont {C.}~\bibnamefont
  {Czaplewski}}, \bibinfo {author} {\bibfnamefont {A.}~\bibnamefont {Liwo}},
  \bibinfo {author} {\bibfnamefont {J.}~\bibnamefont {Lee}}, \bibinfo {author}
  {\bibfnamefont {D.~R.}\ \bibnamefont {Ripoll}}, \bibinfo {author}
  {\bibfnamefont {R.}~\bibnamefont {Kaźmierkiewicz}}, \bibinfo {author}
  {\bibfnamefont {S.}~\bibnamefont {Ołdziej}}, \bibinfo {author}
  {\bibfnamefont {W.~J.}\ \bibnamefont {Wedemeyer}}, \bibinfo {author}
  {\bibfnamefont {K.~D.}\ \bibnamefont {Gibson}}, \bibinfo {author}
  {\bibfnamefont {Y.~A.}\ \bibnamefont {Arnautova}}, \bibinfo {author}
  {\bibfnamefont {J.}~\bibnamefont {Saunders}}, \bibinfo {author}
  {\bibfnamefont {Y.-J.}\ \bibnamefont {Ye}}, \ and\ \bibinfo {author}
  {\bibfnamefont {H.~A.}\ \bibnamefont {Scheraga}},\ }\bibfield  {title}
  {\enquote {\bibinfo {title} {Recent improvements in prediction of protein
  structure by global optimization of a potential energy function},}\ }\href
  {\doibase 10.1073/pnas.041609598} {\bibfield  {journal} {\bibinfo  {journal}
  {PNAS}\ }\textbf {\bibinfo {volume} {98}},\ \bibinfo {pages} {2329--2333}
  (\bibinfo {year} {2001})}\BibitemShut {NoStop}%
\bibitem [{\citenamefont {Wales}\ and\ \citenamefont {Scheraga}(1999)}]{P4232}%
  \BibitemOpen
  \bibfield  {author} {\bibinfo {author} {\bibfnamefont {D.~J.}\ \bibnamefont
  {Wales}}\ and\ \bibinfo {author} {\bibfnamefont {H.~A.}\ \bibnamefont
  {Scheraga}},\ }\bibfield  {title} {\enquote {\bibinfo {title} {Global
  optimization of clusters, crystals, and biomolecules},}\ }\href@noop {}
  {\bibfield  {journal} {\bibinfo  {journal} {Science}\ }\textbf {\bibinfo
  {volume} {285}},\ \bibinfo {pages} {1368} (\bibinfo {year}
  {1999})}\BibitemShut {NoStop}%
\bibitem [{\citenamefont {Floudas}\ and\ \citenamefont
  {Gounaris}(2009)}]{P5606}%
  \BibitemOpen
  \bibfield  {author} {\bibinfo {author} {\bibfnamefont {C.~A.}\ \bibnamefont
  {Floudas}}\ and\ \bibinfo {author} {\bibfnamefont {C.~E.}\ \bibnamefont
  {Gounaris}},\ }\bibfield  {title} {\enquote {\bibinfo {title} {A review of
  recent advances in global optimization},}\ }\href@noop {} {\bibfield
  {journal} {\bibinfo  {journal} {J. Glob. Optim.}\ }\textbf {\bibinfo {volume}
  {45}},\ \bibinfo {pages} {3--38} (\bibinfo {year} {2009})}\BibitemShut
  {NoStop}%
\bibitem [{\citenamefont {Hartke}(2011)}]{hartke_global_2011}%
  \BibitemOpen
  \bibfield  {author} {\bibinfo {author} {\bibfnamefont {B.}~\bibnamefont
  {Hartke}},\ }\bibfield  {title} {\enquote {\bibinfo {title} {Global
  optimization},}\ }\href {\doibase 10.1002/wcms.70} {\bibfield  {journal}
  {\bibinfo  {journal} {Wiley Interdisciplinary Reviews: Computational
  Molecular Science}\ }\textbf {\bibinfo {volume} {1}},\ \bibinfo {pages}
  {879--887} (\bibinfo {year} {2011})}\BibitemShut {NoStop}%
\bibitem [{\citenamefont {Bone}\ and\ \citenamefont
  {Villar}(1997)}]{bone_exhaustive_1997}%
  \BibitemOpen
  \bibfield  {author} {\bibinfo {author} {\bibfnamefont {R.~G.~A.}\
  \bibnamefont {Bone}}\ and\ \bibinfo {author} {\bibfnamefont {H.~O.}\
  \bibnamefont {Villar}},\ }\bibfield  {title} {\enquote {\bibinfo {title}
  {Exhaustive enumeration of molecular substructures},}\ }\href {\doibase
  10.1002/(SICI)1096-987X(19970115)18:1<86::AID-JCC9>3.0.CO;2-W} {\bibfield
  {journal} {\bibinfo  {journal} {Journal of Computational Chemistry}\ }\textbf
  {\bibinfo {volume} {18}},\ \bibinfo {pages} {86--107} (\bibinfo {year}
  {1997})}\BibitemShut {NoStop}%
\bibitem [{\citenamefont {Cyvin}\ \emph {et~al.}(1997)\citenamefont {Cyvin},
  \citenamefont {Wang}, \citenamefont {Brunvoll}, \citenamefont {Cao},
  \citenamefont {Li}, \citenamefont {Cyvin},\ and\ \citenamefont
  {Wang}}]{cyvin_staggered_1997}%
  \BibitemOpen
  \bibfield  {author} {\bibinfo {author} {\bibfnamefont {S.~J.}\ \bibnamefont
  {Cyvin}}, \bibinfo {author} {\bibfnamefont {J.}~\bibnamefont {Wang}},
  \bibinfo {author} {\bibfnamefont {J.}~\bibnamefont {Brunvoll}}, \bibinfo
  {author} {\bibfnamefont {S.}~\bibnamefont {Cao}}, \bibinfo {author}
  {\bibfnamefont {Y.}~\bibnamefont {Li}}, \bibinfo {author} {\bibfnamefont
  {B.~N.}\ \bibnamefont {Cyvin}}, \ and\ \bibinfo {author} {\bibfnamefont
  {Y.}~\bibnamefont {Wang}},\ }\bibfield  {title} {\enquote {\bibinfo {title}
  {Staggered conformers of alkanes: complete solution of the enumeration
  problem},}\ }\href {\doibase 10.1016/S0022-2860(97)00025-2} {\bibfield
  {journal} {\bibinfo  {journal} {Journal of Molecular Structure}\ }\bibinfo
  {series} {Structural {Chemistry}},\ \textbf {\bibinfo {volume} {413-414}},\
  \bibinfo {pages} {227--239} (\bibinfo {year} {1997})}\BibitemShut {NoStop}%
\bibitem [{\citenamefont {Hu}\ and\ \citenamefont
  {Kuhlman}(2006)}]{hu_protein_2006}%
  \BibitemOpen
  \bibfield  {author} {\bibinfo {author} {\bibfnamefont {X.}~\bibnamefont
  {Hu}}\ and\ \bibinfo {author} {\bibfnamefont {B.}~\bibnamefont {Kuhlman}},\
  }\bibfield  {title} {\enquote {\bibinfo {title} {Protein design simulations
  suggest that side-chain conformational entropy is not a strong determinant of
  amino acid environmental preferences},}\ }\href {\doibase 10.1002/prot.20786}
  {\bibfield  {journal} {\bibinfo  {journal} {Proteins: Structure, Function,
  and Bioinformatics}\ }\textbf {\bibinfo {volume} {62}},\ \bibinfo {pages}
  {739--748} (\bibinfo {year} {2006})}\BibitemShut {NoStop}%
\bibitem [{\citenamefont {Pollock}\ \emph {et~al.}(2008)\citenamefont
  {Pollock}, \citenamefont {Coutsias}, \citenamefont {Wester},\ and\
  \citenamefont {Oprea}}]{pollock_scaffold_2008}%
  \BibitemOpen
  \bibfield  {author} {\bibinfo {author} {\bibfnamefont {S.~N.}\ \bibnamefont
  {Pollock}}, \bibinfo {author} {\bibfnamefont {E.~A.}\ \bibnamefont
  {Coutsias}}, \bibinfo {author} {\bibfnamefont {M.~J.}\ \bibnamefont
  {Wester}}, \ and\ \bibinfo {author} {\bibfnamefont {T.~I.}\ \bibnamefont
  {Oprea}},\ }\bibfield  {title} {\enquote {\bibinfo {title} {Scaffold
  {Topologies}. 1. {Exhaustive} {Enumeration} up to {Eight} {Rings}},}\ }\href
  {\doibase 10.1021/ci7003412} {\bibfield  {journal} {\bibinfo  {journal} {J.
  Chem. Inf. Model.}\ }\textbf {\bibinfo {volume} {48}},\ \bibinfo {pages}
  {1304--1310} (\bibinfo {year} {2008})}\BibitemShut {NoStop}%
\bibitem [{\citenamefont {Rahm}\ and\ \citenamefont
  {Erhart}(2017)}]{rahm_beyond_2017}%
  \BibitemOpen
  \bibfield  {author} {\bibinfo {author} {\bibfnamefont {J.~M.}\ \bibnamefont
  {Rahm}}\ and\ \bibinfo {author} {\bibfnamefont {P.}~\bibnamefont {Erhart}},\
  }\bibfield  {title} {\enquote {\bibinfo {title} {Beyond {Magic} {Numbers}:
  {Atomic} {Scale} {Equilibrium} {Nanoparticle} {Shapes} for {Any} {Size}},}\
  }\href {\doibase 10.1021/acs.nanolett.7b02761} {\bibfield  {journal}
  {\bibinfo  {journal} {Nano Lett.}\ }\textbf {\bibinfo {volume} {17}},\
  \bibinfo {pages} {5775--5781} (\bibinfo {year} {2017})}\BibitemShut {NoStop}%
\bibitem [{\citenamefont {Metropolis}\ \emph {et~al.}(1953)\citenamefont
  {Metropolis}, \citenamefont {Rosenbluth}, \citenamefont {Rosenbluth},
  \citenamefont {Teller},\ and\ \citenamefont
  {Teller}}]{metropolis_equation_1953}%
  \BibitemOpen
  \bibfield  {author} {\bibinfo {author} {\bibfnamefont {N.}~\bibnamefont
  {Metropolis}}, \bibinfo {author} {\bibfnamefont {A.~W.}\ \bibnamefont
  {Rosenbluth}}, \bibinfo {author} {\bibfnamefont {M.~N.}\ \bibnamefont
  {Rosenbluth}}, \bibinfo {author} {\bibfnamefont {A.~H.}\ \bibnamefont
  {Teller}}, \ and\ \bibinfo {author} {\bibfnamefont {E.}~\bibnamefont
  {Teller}},\ }\bibfield  {title} {\enquote {\bibinfo {title} {Equation of
  {State} {Calculations} by {Fast} {Computing} {Machines}},}\ }\href {\doibase
  10.1063/1.1699114} {\bibfield  {journal} {\bibinfo  {journal} {J. Chem.
  Phys.}\ }\textbf {\bibinfo {volume} {21}},\ \bibinfo {pages} {1087--1092}
  (\bibinfo {year} {1953})}\BibitemShut {NoStop}%
\bibitem [{\citenamefont {Laks}\ \emph {et~al.}(1992)\citenamefont {Laks},
  \citenamefont {Ferreira}, \citenamefont {Froyen},\ and\ \citenamefont
  {Zunger}}]{P5557}%
  \BibitemOpen
  \bibfield  {author} {\bibinfo {author} {\bibfnamefont {D.~B.}\ \bibnamefont
  {Laks}}, \bibinfo {author} {\bibfnamefont {L.~G.}\ \bibnamefont {Ferreira}},
  \bibinfo {author} {\bibfnamefont {S.}~\bibnamefont {Froyen}}, \ and\ \bibinfo
  {author} {\bibfnamefont {A.}~\bibnamefont {Zunger}},\ }\bibfield  {title}
  {\enquote {\bibinfo {title} {Efficient cluster expansion for substitutional
  systems},}\ }\href@noop {} {\bibfield  {journal} {\bibinfo  {journal} {Phys.
  Rev. B}\ }\textbf {\bibinfo {volume} {46}},\ \bibinfo {pages} {12587}
  (\bibinfo {year} {1992})}\BibitemShut {NoStop}%
\bibitem [{\citenamefont {Shao}, \citenamefont {Cheng},\ and\ \citenamefont
  {Cai}(2004)}]{shao_dynamic_2004}%
  \BibitemOpen
  \bibfield  {author} {\bibinfo {author} {\bibfnamefont {X.}~\bibnamefont
  {Shao}}, \bibinfo {author} {\bibfnamefont {L.}~\bibnamefont {Cheng}}, \ and\
  \bibinfo {author} {\bibfnamefont {W.}~\bibnamefont {Cai}},\ }\bibfield
  {title} {\enquote {\bibinfo {title} {A dynamic lattice searching method for
  fast optimization of {Lennard}–{Jones} clusters},}\ }\href {\doibase
  10.1002/jcc.20096} {\bibfield  {journal} {\bibinfo  {journal} {Journal of
  Computational Chemistry}\ }\textbf {\bibinfo {volume} {25}},\ \bibinfo
  {pages} {1693--1698} (\bibinfo {year} {2004})}\BibitemShut {NoStop}%
\bibitem [{\citenamefont {Yu}\ \emph {et~al.}(2019)\citenamefont {Yu},
  \citenamefont {Wang}, \citenamefont {Chen},\ and\ \citenamefont
  {Wang}}]{yu_unbiased_2019}%
  \BibitemOpen
  \bibfield  {author} {\bibinfo {author} {\bibfnamefont {K.}~\bibnamefont
  {Yu}}, \bibinfo {author} {\bibfnamefont {X.}~\bibnamefont {Wang}}, \bibinfo
  {author} {\bibfnamefont {L.}~\bibnamefont {Chen}}, \ and\ \bibinfo {author}
  {\bibfnamefont {L.}~\bibnamefont {Wang}},\ }\bibfield  {title} {\enquote
  {\bibinfo {title} {Unbiased fuzzy global optimization of {Lennard}-{Jones}
  clusters for n <= 1000},}\ }\href {\doibase 10.1063/1.5127913} {\bibfield
  {journal} {\bibinfo  {journal} {J. Chem. Phys.}\ }\textbf {\bibinfo {volume}
  {151}},\ \bibinfo {pages} {214105} (\bibinfo {year} {2019})}\BibitemShut
  {NoStop}%
\bibitem [{\citenamefont {Wales}\ and\ \citenamefont
  {Doye}(1997)}]{wales_global_1997}%
  \BibitemOpen
  \bibfield  {author} {\bibinfo {author} {\bibfnamefont {D.~J.}\ \bibnamefont
  {Wales}}\ and\ \bibinfo {author} {\bibfnamefont {J.~P.~K.}\ \bibnamefont
  {Doye}},\ }\bibfield  {title} {\enquote {\bibinfo {title} {Global
  {Optimization} by {Basin}-{Hopping} and the {Lowest} {Energy} {Structures} of
  {Lennard}-{Jones} {Clusters} {Containing} up to 110 {Atoms}},}\ }\href
  {\doibase 10.1021/jp970984n} {\bibfield  {journal} {\bibinfo  {journal} {J.
  Phys. Chem. A}\ }\textbf {\bibinfo {volume} {101}},\ \bibinfo {pages}
  {5111--5116} (\bibinfo {year} {1997})}\BibitemShut {NoStop}%
\bibitem [{\citenamefont {Kirkpatrick}, \citenamefont {Gelatt},\ and\
  \citenamefont {Vecchi}(1983)}]{kirkpatrick_optimization_1983}%
  \BibitemOpen
  \bibfield  {author} {\bibinfo {author} {\bibfnamefont {S.}~\bibnamefont
  {Kirkpatrick}}, \bibinfo {author} {\bibfnamefont {C.~D.}\ \bibnamefont
  {Gelatt}}, \ and\ \bibinfo {author} {\bibfnamefont {M.~P.}\ \bibnamefont
  {Vecchi}},\ }\bibfield  {title} {\enquote {\bibinfo {title} {Optimization by
  {Simulated} {Annealing}},}\ }\href {\doibase 10.1126/science.220.4598.671}
  {\bibfield  {journal} {\bibinfo  {journal} {Science}\ }\textbf {\bibinfo
  {volume} {220}},\ \bibinfo {pages} {671--680} (\bibinfo {year}
  {1983})}\BibitemShut {NoStop}%
\bibitem [{\citenamefont {Goedecker}(2004)}]{goedecker_minima_2004}%
  \BibitemOpen
  \bibfield  {author} {\bibinfo {author} {\bibfnamefont {S.}~\bibnamefont
  {Goedecker}},\ }\bibfield  {title} {\enquote {\bibinfo {title} {Minima
  hopping: {An} efficient search method for the global minimum of the potential
  energy surface of complex molecular systems},}\ }\href {\doibase
  10.1063/1.1724816} {\bibfield  {journal} {\bibinfo  {journal} {J. Chem.
  Phys.}\ }\textbf {\bibinfo {volume} {120}},\ \bibinfo {pages} {9911--9917}
  (\bibinfo {year} {2004})}\BibitemShut {NoStop}%
\bibitem [{\citenamefont {Schoenborn}\ \emph {et~al.}(2009)\citenamefont
  {Schoenborn}, \citenamefont {Goedecker}, \citenamefont {Roy},\ and\
  \citenamefont {Oganov}}]{P2201}%
  \BibitemOpen
  \bibfield  {author} {\bibinfo {author} {\bibfnamefont {S.~E.}\ \bibnamefont
  {Schoenborn}}, \bibinfo {author} {\bibfnamefont {S.}~\bibnamefont
  {Goedecker}}, \bibinfo {author} {\bibfnamefont {S.}~\bibnamefont {Roy}}, \
  and\ \bibinfo {author} {\bibfnamefont {A.~R.}\ \bibnamefont {Oganov}},\
  }\bibfield  {title} {\enquote {\bibinfo {title} {The performance of minima
  hopping and evolutionary algorithms for cluster structure prediction},}\
  }\href@noop {} {\bibfield  {journal} {\bibinfo  {journal} {J. Chem. Phys.}\
  }\textbf {\bibinfo {volume} {130}},\ \bibinfo {pages} {144108} (\bibinfo
  {year} {2009})}\BibitemShut {NoStop}%
\bibitem [{\citenamefont {Z.~Michalewicz}(1991)}]{P1518}%
  \BibitemOpen
  \bibfield  {author} {\bibinfo {author} {\bibfnamefont {C.~J.}\ \bibnamefont
  {Z.~Michalewicz}},\ }\bibfield  {title} {\enquote {\bibinfo {title} {Genetic
  algorithms for numerical optimization},}\ }\href@noop {} {\bibfield
  {journal} {\bibinfo  {journal} {Statistics and Computing}\ }\textbf {\bibinfo
  {volume} {1}},\ \bibinfo {pages} {75} (\bibinfo {year} {1991})}\BibitemShut
  {NoStop}%
\bibitem [{\citenamefont {Deaven}\ and\ \citenamefont
  {Ho}(1995)}]{deaven_molecular_1995}%
  \BibitemOpen
  \bibfield  {author} {\bibinfo {author} {\bibfnamefont {D.~M.}\ \bibnamefont
  {Deaven}}\ and\ \bibinfo {author} {\bibfnamefont {K.~M.}\ \bibnamefont
  {Ho}},\ }\bibfield  {title} {\enquote {\bibinfo {title} {Molecular {Geometry}
  {Optimization} with a {Genetic} {Algorithm}},}\ }\href {\doibase
  10.1103/PhysRevLett.75.288} {\bibfield  {journal} {\bibinfo  {journal} {Phys.
  Rev. Lett.}\ }\textbf {\bibinfo {volume} {75}},\ \bibinfo {pages} {288--291}
  (\bibinfo {year} {1995})}\BibitemShut {NoStop}%
\bibitem [{\citenamefont {Oganov}\ and\ \citenamefont {Glass}(2006)}]{P0862}%
  \BibitemOpen
  \bibfield  {author} {\bibinfo {author} {\bibfnamefont {A.~R.}\ \bibnamefont
  {Oganov}}\ and\ \bibinfo {author} {\bibfnamefont {C.~W.}\ \bibnamefont
  {Glass}},\ }\bibfield  {title} {\enquote {\bibinfo {title} {Crystal structure
  predicition using ab initio evolutionary techniques: Principles and
  applications},}\ }\href@noop {} {\bibfield  {journal} {\bibinfo  {journal}
  {J. Chem. Phys.}\ }\textbf {\bibinfo {volume} {124}},\ \bibinfo {pages}
  {244704} (\bibinfo {year} {2006})}\BibitemShut {NoStop}%
\bibitem [{\citenamefont {Vilhelmsen}\ and\ \citenamefont
  {Hammer}(2012)}]{vilhelmsen_systematic_2012}%
  \BibitemOpen
  \bibfield  {author} {\bibinfo {author} {\bibfnamefont {L.~B.}\ \bibnamefont
  {Vilhelmsen}}\ and\ \bibinfo {author} {\bibfnamefont {B.}~\bibnamefont
  {Hammer}},\ }\bibfield  {title} {\enquote {\bibinfo {title} {Systematic
  {Study} of \$\{{\textbackslash}mathrm\{{Au}\}\}\_\{6\}\$ to
  \$\{{\textbackslash}mathrm\{{Au}\}\}\_\{12\}\$ {Gold} {Clusters} on
  {MgO}(100) \${F}\$ {Centers} {Using} {Density}-{Functional} {Theory}},}\
  }\href {\doibase 10.1103/PhysRevLett.108.126101} {\bibfield  {journal}
  {\bibinfo  {journal} {Phys. Rev. Lett.}\ }\textbf {\bibinfo {volume} {108}},\
  \bibinfo {pages} {126101} (\bibinfo {year} {2012})}\BibitemShut {NoStop}%
\bibitem [{\citenamefont {Vilhelmsen}\ and\ \citenamefont
  {Hammer}(2014{\natexlab{a}})}]{vilhelmsen_identification_2014}%
  \BibitemOpen
  \bibfield  {author} {\bibinfo {author} {\bibfnamefont {L.~B.}\ \bibnamefont
  {Vilhelmsen}}\ and\ \bibinfo {author} {\bibfnamefont {B.}~\bibnamefont
  {Hammer}},\ }\bibfield  {title} {\enquote {\bibinfo {title} {Identification
  of the {Catalytic} {Site} at the {Interface} {Perimeter} of {Au} {Clusters}
  on {Rutile} {TiO}2(110)},}\ }\href {\doibase 10.1021/cs500202f} {\bibfield
  {journal} {\bibinfo  {journal} {ACS Catal.}\ }\textbf {\bibinfo {volume}
  {4}},\ \bibinfo {pages} {1626--1631} (\bibinfo {year}
  {2014}{\natexlab{a}})}\BibitemShut {NoStop}%
\bibitem [{\citenamefont {Huang}\ \emph {et~al.}(2019)\citenamefont {Huang},
  \citenamefont {Jiang}, \citenamefont {Liang}, \citenamefont {Wu},
  \citenamefont {Li},\ and\ \citenamefont {Hou}}]{huang_structural_2019}%
  \BibitemOpen
  \bibfield  {author} {\bibinfo {author} {\bibfnamefont {P.}~\bibnamefont
  {Huang}}, \bibinfo {author} {\bibfnamefont {Y.}~\bibnamefont {Jiang}},
  \bibinfo {author} {\bibfnamefont {T.}~\bibnamefont {Liang}}, \bibinfo
  {author} {\bibfnamefont {E.}~\bibnamefont {Wu}}, \bibinfo {author}
  {\bibfnamefont {J.}~\bibnamefont {Li}}, \ and\ \bibinfo {author}
  {\bibfnamefont {J.}~\bibnamefont {Hou}},\ }\bibfield  {title} {\enquote
  {\bibinfo {title} {Structural exploration of auxm- (m = si, ge, sn; x = 9-12)
  clusters with a revised genetic algorithm},}\ }\href {\doibase
  10.1039/C9RA01019J} {\bibfield  {journal} {\bibinfo  {journal} {RSC Adv.}\
  }\textbf {\bibinfo {volume} {9}},\ \bibinfo {pages} {7432--7439} (\bibinfo
  {year} {2019})}\BibitemShut {NoStop}%
\bibitem [{\citenamefont {Buendía}\ \emph {et~al.}(2017)\citenamefont
  {Buendía}, \citenamefont {Vargas}, \citenamefont {Johnston},\ and\
  \citenamefont {Beltrán}}]{buendia_study_2017}%
  \BibitemOpen
  \bibfield  {author} {\bibinfo {author} {\bibfnamefont {F.}~\bibnamefont
  {Buendía}}, \bibinfo {author} {\bibfnamefont {J.~A.}\ \bibnamefont
  {Vargas}}, \bibinfo {author} {\bibfnamefont {R.~L.}\ \bibnamefont
  {Johnston}}, \ and\ \bibinfo {author} {\bibfnamefont {M.~R.}\ \bibnamefont
  {Beltrán}},\ }\bibfield  {title} {\enquote {\bibinfo {title} {Study of the
  stability of small {AuRh} clusters found by a {Genetic} {Algorithm}
  methodology},}\ }\href {\doibase 10.1016/j.comptc.2017.09.008} {\bibfield
  {journal} {\bibinfo  {journal} {Computational and Theoretical Chemistry}\
  }\textbf {\bibinfo {volume} {1119}},\ \bibinfo {pages} {51--58} (\bibinfo
  {year} {2017})}\BibitemShut {NoStop}%
\bibitem [{\citenamefont {Heydariyan}\ \emph {et~al.}(2018)\citenamefont
  {Heydariyan}, \citenamefont {Nouri}, \citenamefont {Alaei}, \citenamefont
  {Allahyari},\ and\ \citenamefont {Niehaus}}]{heydariyan_new_2018}%
  \BibitemOpen
  \bibfield  {author} {\bibinfo {author} {\bibfnamefont {S.}~\bibnamefont
  {Heydariyan}}, \bibinfo {author} {\bibfnamefont {M.~R.}\ \bibnamefont
  {Nouri}}, \bibinfo {author} {\bibfnamefont {M.}~\bibnamefont {Alaei}},
  \bibinfo {author} {\bibfnamefont {Z.}~\bibnamefont {Allahyari}}, \ and\
  \bibinfo {author} {\bibfnamefont {T.~A.}\ \bibnamefont {Niehaus}},\
  }\bibfield  {title} {\enquote {\bibinfo {title} {New candidates for the
  global minimum of medium-sized silicon clusters: {A} hybrid {DFTB}/{DFT}
  genetic algorithm applied to {Sin}, n = 8-80},}\ }\href {\doibase
  10.1063/1.5037159} {\bibfield  {journal} {\bibinfo  {journal} {J. Chem.
  Phys.}\ }\textbf {\bibinfo {volume} {149}},\ \bibinfo {pages} {074313}
  (\bibinfo {year} {2018})}\BibitemShut {NoStop}%
\bibitem [{\citenamefont {Bozkurt}\ \emph {et~al.}(2018)\citenamefont
  {Bozkurt}, \citenamefont {Perez}, \citenamefont {Hovius}, \citenamefont
  {Browning},\ and\ \citenamefont {Rothlisberger}}]{bozkurt_genetic_2018}%
  \BibitemOpen
  \bibfield  {author} {\bibinfo {author} {\bibfnamefont {E.}~\bibnamefont
  {Bozkurt}}, \bibinfo {author} {\bibfnamefont {M.~A.~S.}\ \bibnamefont
  {Perez}}, \bibinfo {author} {\bibfnamefont {R.}~\bibnamefont {Hovius}},
  \bibinfo {author} {\bibfnamefont {N.~J.}\ \bibnamefont {Browning}}, \ and\
  \bibinfo {author} {\bibfnamefont {U.}~\bibnamefont {Rothlisberger}},\
  }\bibfield  {title} {\enquote {\bibinfo {title} {Genetic {Algorithm} {Based}
  {Design} and {Experimental} {Characterization} of a {Highly} {Thermostable}
  {Metalloprotein}},}\ }\href {\doibase 10.1021/jacs.7b10660} {\bibfield
  {journal} {\bibinfo  {journal} {J. Am. Chem. Soc.}\ }\textbf {\bibinfo
  {volume} {140}},\ \bibinfo {pages} {4517--4521} (\bibinfo {year}
  {2018})}\BibitemShut {NoStop}%
\bibitem [{\citenamefont {Rondina}\ and\ \citenamefont {{Da
  Silva}}(2013)}]{P3937}%
  \BibitemOpen
  \bibfield  {author} {\bibinfo {author} {\bibfnamefont {G.~G.}\ \bibnamefont
  {Rondina}}\ and\ \bibinfo {author} {\bibfnamefont {J.~L.~F.}\ \bibnamefont
  {{Da Silva}}},\ }\bibfield  {title} {\enquote {\bibinfo {title} {Revised
  basin-hopping monte carlo algorithm for structure optimization of clusters
  and nanoparticles},}\ }\href@noop {} {\bibfield  {journal} {\bibinfo
  {journal} {J. Chem. Inf. Mod.}\ }\textbf {\bibinfo {volume} {53}},\ \bibinfo
  {pages} {2282} (\bibinfo {year} {2013})}\BibitemShut {NoStop}%
\bibitem [{\citenamefont {{Jones J. E.}}\ and\ \citenamefont {{Chapman
  Sydney}}(1924)}]{jones_j._e._determination_1924}%
  \BibitemOpen
  \bibfield  {author} {\bibinfo {author} {\bibnamefont {{Jones J. E.}}}\ and\
  \bibinfo {author} {\bibnamefont {{Chapman Sydney}}},\ }\bibfield  {title}
  {\enquote {\bibinfo {title} {On the determination of molecular fields.
  —{II}. {From} the equation of state of a gas},}\ }\href {\doibase
  10.1098/rspa.1924.0082} {\bibfield  {journal} {\bibinfo  {journal}
  {Proceedings of the Royal Society of London. Series A, Containing Papers of a
  Mathematical and Physical Character}\ }\textbf {\bibinfo {volume} {106}},\
  \bibinfo {pages} {463--477} (\bibinfo {year} {1924})}\BibitemShut {NoStop}%
\bibitem [{\citenamefont {Eshet}, \citenamefont {Bruneval},\ and\ \citenamefont
  {Parrinello}(2008)}]{P1978}%
  \BibitemOpen
  \bibfield  {author} {\bibinfo {author} {\bibfnamefont {H.}~\bibnamefont
  {Eshet}}, \bibinfo {author} {\bibfnamefont {F.}~\bibnamefont {Bruneval}}, \
  and\ \bibinfo {author} {\bibfnamefont {M.}~\bibnamefont {Parrinello}},\
  }\bibfield  {title} {\enquote {\bibinfo {title} {New lennard-jones metastable
  phase},}\ }\href@noop {} {\bibfield  {journal} {\bibinfo  {journal} {J. Chem.
  Phys.}\ }\textbf {\bibinfo {volume} {129}},\ \bibinfo {pages} {026101}
  (\bibinfo {year} {2008})}\BibitemShut {NoStop}%
\bibitem [{\citenamefont {Li}\ and\ \citenamefont
  {Scheraga}(1987)}]{li_monte_1987}%
  \BibitemOpen
  \bibfield  {author} {\bibinfo {author} {\bibfnamefont {Z.}~\bibnamefont
  {Li}}\ and\ \bibinfo {author} {\bibfnamefont {H.~A.}\ \bibnamefont
  {Scheraga}},\ }\bibfield  {title} {\enquote {\bibinfo {title} {Monte
  {Carlo}-minimization approach to the multiple-minima problem in protein
  folding},}\ }\href {\doibase 10.1073/pnas.84.19.6611} {\bibfield  {journal}
  {\bibinfo  {journal} {PNAS}\ }\textbf {\bibinfo {volume} {84}},\ \bibinfo
  {pages} {6611--6615} (\bibinfo {year} {1987})}\BibitemShut {NoStop}%
\bibitem [{\citenamefont {Frenkel}\ and\ \citenamefont
  {Smit}(2002)}]{frenkel_daan_understanding_2002}%
  \BibitemOpen
  \bibfield  {author} {\bibinfo {author} {\bibfnamefont {D.}~\bibnamefont
  {Frenkel}}\ and\ \bibinfo {author} {\bibfnamefont {B.}~\bibnamefont {Smit}},\
  }\href@noop {} {\emph {\bibinfo {title} {Understanding {Molecular}
  {Simulations}}}}\ (\bibinfo  {publisher} {Academic Press},\ \bibinfo {year}
  {2002})\BibitemShut {NoStop}%
\bibitem [{\citenamefont {Takeuchi}(2006)}]{takeuchi_clever_2006}%
  \BibitemOpen
  \bibfield  {author} {\bibinfo {author} {\bibfnamefont {H.}~\bibnamefont
  {Takeuchi}},\ }\bibfield  {title} {\enquote {\bibinfo {title} {Clever and
  {Efficient} {Method} for {Searching} {Optimal} {Geometries} of
  {Lennard}-{Jones} {Clusters}},}\ }\href {\doibase 10.1021/ci600206k}
  {\bibfield  {journal} {\bibinfo  {journal} {J. Chem. Inf. Model.}\ }\textbf
  {\bibinfo {volume} {46}},\ \bibinfo {pages} {2066--2070} (\bibinfo {year}
  {2006})}\BibitemShut {NoStop}%
\bibitem [{\citenamefont {Hunter}(2007)}]{hunter_matplotlib_2007}%
  \BibitemOpen
  \bibfield  {author} {\bibinfo {author} {\bibfnamefont {J.~D.}\ \bibnamefont
  {Hunter}},\ }\bibfield  {title} {\enquote {\bibinfo {title} {Matplotlib {A}
  2d {Graphics} {Environment}},}\ }\href {\doibase 10.1109/MCSE.2007.55}
  {\bibfield  {journal} {\bibinfo  {journal} {Computing in Science
  Engineering}\ }\textbf {\bibinfo {volume} {9}},\ \bibinfo {pages} {90--95}
  (\bibinfo {year} {2007})}\BibitemShut {NoStop}%
\bibitem [{\citenamefont {M\"uller}\ and\ \citenamefont
  {Zunger}(2001)}]{P2119}%
  \BibitemOpen
  \bibfield  {author} {\bibinfo {author} {\bibfnamefont {S.}~\bibnamefont
  {M\"uller}}\ and\ \bibinfo {author} {\bibfnamefont {A.}~\bibnamefont
  {Zunger}},\ }\bibfield  {title} {\enquote {\bibinfo {title} {Structure of
  ordered and disordered $\alpha$-brass},}\ }\href@noop {} {\bibfield
  {journal} {\bibinfo  {journal} {Phys. Rev. B}\ }\textbf {\bibinfo {volume}
  {63}},\ \bibinfo {pages} {094204} (\bibinfo {year} {2001})}\BibitemShut
  {NoStop}%
\bibitem [{\citenamefont {Stukowski}(2009)}]{stukowski_visualization_2009}%
  \BibitemOpen
  \bibfield  {author} {\bibinfo {author} {\bibfnamefont {A.}~\bibnamefont
  {Stukowski}},\ }\bibfield  {title} {\enquote {\bibinfo {title} {Visualization
  and analysis of atomistic simulation data with {OVITO}–the {Open}
  {Visualization} {Tool}},}\ }\href {\doibase 10.1088/0965-0393/18/1/015012}
  {\bibfield  {journal} {\bibinfo  {journal} {Modelling Simul. Mater. Sci.
  Eng.}\ }\textbf {\bibinfo {volume} {18}},\ \bibinfo {pages} {015012}
  (\bibinfo {year} {2009})}\BibitemShut {NoStop}%
\bibitem [{\citenamefont {Vilhelmsen}\ and\ \citenamefont
  {Hammer}(2014{\natexlab{b}})}]{vilhelmsen_genetic_2014}%
  \BibitemOpen
  \bibfield  {author} {\bibinfo {author} {\bibfnamefont {L.~B.}\ \bibnamefont
  {Vilhelmsen}}\ and\ \bibinfo {author} {\bibfnamefont {B.}~\bibnamefont
  {Hammer}},\ }\bibfield  {title} {\enquote {\bibinfo {title} {A genetic
  algorithm for first principles global structure optimization of supported
  nano structures},}\ }\href {\doibase 10.1063/1.4886337} {\bibfield  {journal}
  {\bibinfo  {journal} {J. Chem. Phys.}\ }\textbf {\bibinfo {volume} {141}},\
  \bibinfo {pages} {044711} (\bibinfo {year} {2014}{\natexlab{b}})}\BibitemShut
  {NoStop}%
\bibitem [{\citenamefont {Plimpton}(1995)}]{P4473}%
  \BibitemOpen
  \bibfield  {author} {\bibinfo {author} {\bibfnamefont {S.}~\bibnamefont
  {Plimpton}},\ }\bibfield  {title} {\enquote {\bibinfo {title} {Fast parallel
  algorithms for short-range molecular dynamics},}\ }\href@noop {} {\bibfield
  {journal} {\bibinfo  {journal} {J. Comp. Phys.}\ }\textbf {\bibinfo {volume}
  {117}},\ \bibinfo {pages} {1} (\bibinfo {year} {1995})}\BibitemShut {NoStop}%
\bibitem [{noa({\natexlab{a}})}]{noauthor_gpl_nodate}%
  \BibitemOpen
  \href {https://www.gnu.org/licenses/gpl-3.0.en.html} {\enquote {\bibinfo
  {title} {{GPL} v3},}\ } ({\natexlab{a}}),\ \bibinfo {note}
  {https://www.gnu.org/licenses/gpl-3.0.en.html}\BibitemShut {NoStop}%
\bibitem [{noa({\natexlab{b}})}]{noauthor_cambridge_nodate}%
  \BibitemOpen
  \href {http://www-wales.ch.cam.ac.uk/CCD.html} {\enquote {\bibinfo {title}
  {Cambridge {Energy} {Landscape} {Database}},}\ } ({\natexlab{b}}),\ \bibinfo
  {note} {\url{http://www-wales.ch.cam.ac.uk/CCD.html}}\BibitemShut {NoStop}%
\bibitem [{\citenamefont {Larsen}\ \emph {et~al.}(2017)\citenamefont {Larsen},
  \citenamefont {Mortensen}, \citenamefont {Blomqvist}, \citenamefont
  {Castelli}, \citenamefont {Christensen}, \citenamefont
  {Du{\textbackslash}lak}, \citenamefont {Friis}, \citenamefont {Groves},
  \citenamefont {Hammer}, \citenamefont {Hargus}, \citenamefont {Hermes},
  \citenamefont {Jennings}, \citenamefont {Jensen}, \citenamefont {Kermode},
  \citenamefont {Kitchin}, \citenamefont {Kolsbjerg}, \citenamefont {Kubal},
  \citenamefont {Kaasbjerg}, \citenamefont {Lysgaard}, \citenamefont
  {Maronsson}, \citenamefont {Maxson}, \citenamefont {Olsen}, \citenamefont
  {Pastewka}, \citenamefont {Peterson}, \citenamefont {Rostgaard},
  \citenamefont {Schiøtz}, \citenamefont {Schütt}, \citenamefont {Strange},
  \citenamefont {Thygesen}, \citenamefont {Vegge}, \citenamefont {Vilhelmsen},
  \citenamefont {Walter}, \citenamefont {Zeng},\ and\ \citenamefont
  {Jacobsen}}]{larsen_atomic_2017}%
  \BibitemOpen
  \bibfield  {author} {\bibinfo {author} {\bibfnamefont {A.~H.}\ \bibnamefont
  {Larsen}}, \bibinfo {author} {\bibfnamefont {J.~J.}\ \bibnamefont
  {Mortensen}}, \bibinfo {author} {\bibfnamefont {J.}~\bibnamefont
  {Blomqvist}}, \bibinfo {author} {\bibfnamefont {I.~E.}\ \bibnamefont
  {Castelli}}, \bibinfo {author} {\bibfnamefont {R.}~\bibnamefont
  {Christensen}}, \bibinfo {author} {\bibfnamefont {M.}~\bibnamefont
  {Du{\textbackslash}lak}}, \bibinfo {author} {\bibfnamefont {J.}~\bibnamefont
  {Friis}}, \bibinfo {author} {\bibfnamefont {M.~N.}\ \bibnamefont {Groves}},
  \bibinfo {author} {\bibfnamefont {B.}~\bibnamefont {Hammer}}, \bibinfo
  {author} {\bibfnamefont {C.}~\bibnamefont {Hargus}}, \bibinfo {author}
  {\bibfnamefont {E.~D.}\ \bibnamefont {Hermes}}, \bibinfo {author}
  {\bibfnamefont {P.~C.}\ \bibnamefont {Jennings}}, \bibinfo {author}
  {\bibfnamefont {P.~B.}\ \bibnamefont {Jensen}}, \bibinfo {author}
  {\bibfnamefont {J.}~\bibnamefont {Kermode}}, \bibinfo {author} {\bibfnamefont
  {J.~R.}\ \bibnamefont {Kitchin}}, \bibinfo {author} {\bibfnamefont {E.~L.}\
  \bibnamefont {Kolsbjerg}}, \bibinfo {author} {\bibfnamefont {J.}~\bibnamefont
  {Kubal}}, \bibinfo {author} {\bibfnamefont {K.}~\bibnamefont {Kaasbjerg}},
  \bibinfo {author} {\bibfnamefont {S.}~\bibnamefont {Lysgaard}}, \bibinfo
  {author} {\bibfnamefont {J.~B.}\ \bibnamefont {Maronsson}}, \bibinfo {author}
  {\bibfnamefont {T.}~\bibnamefont {Maxson}}, \bibinfo {author} {\bibfnamefont
  {T.}~\bibnamefont {Olsen}}, \bibinfo {author} {\bibfnamefont
  {L.}~\bibnamefont {Pastewka}}, \bibinfo {author} {\bibfnamefont
  {A.}~\bibnamefont {Peterson}}, \bibinfo {author} {\bibfnamefont
  {C.}~\bibnamefont {Rostgaard}}, \bibinfo {author} {\bibfnamefont
  {J.}~\bibnamefont {Schiøtz}}, \bibinfo {author} {\bibfnamefont
  {O.}~\bibnamefont {Schütt}}, \bibinfo {author} {\bibfnamefont
  {M.}~\bibnamefont {Strange}}, \bibinfo {author} {\bibfnamefont {K.~S.}\
  \bibnamefont {Thygesen}}, \bibinfo {author} {\bibfnamefont {T.}~\bibnamefont
  {Vegge}}, \bibinfo {author} {\bibfnamefont {L.}~\bibnamefont {Vilhelmsen}},
  \bibinfo {author} {\bibfnamefont {M.}~\bibnamefont {Walter}}, \bibinfo
  {author} {\bibfnamefont {Z.}~\bibnamefont {Zeng}}, \ and\ \bibinfo {author}
  {\bibfnamefont {K.~W.}\ \bibnamefont {Jacobsen}},\ }\bibfield  {title}
  {\enquote {\bibinfo {title} {The atomic simulation environment—a {Python}
  library for working with atoms},}\ }\href {\doibase 10.1088/1361-648X/aa680e}
  {\bibfield  {journal} {\bibinfo  {journal} {J. Phys.: Condens. Matter}\
  }\textbf {\bibinfo {volume} {29}},\ \bibinfo {pages} {273002} (\bibinfo
  {year} {2017})}\BibitemShut {NoStop}%
\bibitem [{\citenamefont {Behler}(2016)}]{P4885}%
  \BibitemOpen
  \bibfield  {author} {\bibinfo {author} {\bibfnamefont {J.}~\bibnamefont
  {Behler}},\ }\bibfield  {title} {\enquote {\bibinfo {title} {Perspective:
  Machine learning potentials for atomistic simulations},}\ }\href@noop {}
  {\bibfield  {journal} {\bibinfo  {journal} {J. Chem. Phys.}\ }\textbf
  {\bibinfo {volume} {145}},\ \bibinfo {pages} {170901} (\bibinfo {year}
  {2016})}\BibitemShut {NoStop}%
\bibitem [{\citenamefont {Rondina}\ and\ \citenamefont
  {Da~Silva}(2013)}]{rondina_revised_2013}%
  \BibitemOpen
  \bibfield  {author} {\bibinfo {author} {\bibfnamefont {G.~G.}\ \bibnamefont
  {Rondina}}\ and\ \bibinfo {author} {\bibfnamefont {J.~L.~F.}\ \bibnamefont
  {Da~Silva}},\ }\bibfield  {title} {\enquote {\bibinfo {title} {Revised
  {Basin}-{Hopping} {Monte} {Carlo} {Algorithm} for {Structure} {Optimization}
  of {Clusters} and {Nanoparticles}},}\ }\href {\doibase 10.1021/ci400224z}
  {\bibfield  {journal} {\bibinfo  {journal} {J. Chem. Inf. Model.}\ }\textbf
  {\bibinfo {volume} {53}},\ \bibinfo {pages} {2282--2298} (\bibinfo {year}
  {2013})}\BibitemShut {NoStop}%
\bibitem [{\citenamefont {Iwamatsu}\ and\ \citenamefont
  {Okabe}(2004)}]{iwamatsu_basin_2004}%
  \BibitemOpen
  \bibfield  {author} {\bibinfo {author} {\bibfnamefont {M.}~\bibnamefont
  {Iwamatsu}}\ and\ \bibinfo {author} {\bibfnamefont {Y.}~\bibnamefont
  {Okabe}},\ }\bibfield  {title} {\enquote {\bibinfo {title} {Basin hopping
  with occasional jumping},}\ }\href {\doibase 10.1016/j.cplett.2004.10.032}
  {\bibfield  {journal} {\bibinfo  {journal} {Chemical Physics Letters}\
  }\textbf {\bibinfo {volume} {399}},\ \bibinfo {pages} {396--400} (\bibinfo
  {year} {2004})}\BibitemShut {NoStop}%
\bibitem [{\citenamefont {Voorhees}(1985)}]{voorhees_theory_1985}%
  \BibitemOpen
  \bibfield  {author} {\bibinfo {author} {\bibfnamefont {P.~W.}\ \bibnamefont
  {Voorhees}},\ }\bibfield  {title} {\enquote {\bibinfo {title} {The theory of
  {Ostwald} ripening},}\ }\href {\doibase 10.1007/BF01017860} {\bibfield
  {journal} {\bibinfo  {journal} {J Stat Phys}\ }\textbf {\bibinfo {volume}
  {38}},\ \bibinfo {pages} {231--252} (\bibinfo {year} {1985})}\BibitemShut
  {NoStop}%
\bibitem [{\citenamefont {Lucas}\ and\ \citenamefont
  {Moskovkin}(2010)}]{lucas_simulation_2010}%
  \BibitemOpen
  \bibfield  {author} {\bibinfo {author} {\bibfnamefont {S.}~\bibnamefont
  {Lucas}}\ and\ \bibinfo {author} {\bibfnamefont {P.}~\bibnamefont
  {Moskovkin}},\ }\bibfield  {title} {\enquote {\bibinfo {title} {Simulation at
  high temperature of atomic deposition, islands coalescence, {Ostwald} and
  inverse {Ostwald} ripening with a general simple kinetic {Monte} {Carlo}
  code},}\ }\href {\doibase 10.1016/j.tsf.2010.04.064} {\bibfield  {journal}
  {\bibinfo  {journal} {Thin Solid Films}\ }\textbf {\bibinfo {volume} {518}},\
  \bibinfo {pages} {5355--5361} (\bibinfo {year} {2010})}\BibitemShut {NoStop}%
\end{thebibliography}%

\end{document}